\documentclass[aps,twocolumn,prl,preprintnumbers,amsmath,amssymb,superscriptaddress]{revtex4-1}

\usepackage{graphicx}
\usepackage{bm}
\usepackage{hyperref}
\usepackage{natbib}
\usepackage{float}

\restylefloat{table}

\begin{document}

\title{Anomalous gap edge dissipation in disordered superconductors on the brink of localization}

\author{Bing Cheng}
\affiliation{Department of Physics and Astronomy, The Johns Hopkins University, Baltimore, MD 21218, USA}

\author{Liang Wu}
\affiliation{Department of Physics and Astronomy, The Johns Hopkins University, Baltimore, MD 21218, USA}

\author{N. J. Laurita}
\affiliation{Department of Physics and Astronomy, The Johns Hopkins University, Baltimore, MD 21218, USA}

\author{Harkirat Singh}
\affiliation{Department of Condensed Matter Physics and Materials Science,
Tata Institute of Fundamental Research, Homi Bhabha Rd., Colaba, Mumbai 400 005, India}

\author{Madhavi Chand}
\affiliation{Department of Condensed Matter Physics and Materials Science,
Tata Institute of Fundamental Research, Homi Bhabha Rd., Colaba, Mumbai 400 005, India}

\author{Pratap Raychaudhuri}
\affiliation{Department of Condensed Matter Physics and Materials Science,
Tata Institute of Fundamental Research, Homi Bhabha Rd., Colaba, Mumbai 400 005, India}

\author{N. P. Armitage}
\affiliation{Department of Physics and Astronomy, The Johns Hopkins University, Baltimore, MD 21218, USA}

\date{\today}

\begin{abstract}

Superconductivity in disordered systems close to an incipient localization transition has been an area of investigation for many years.  It has been noted that in such highly disordered superconductors, anomalous spectral weight develops in their conductivity near and below the superconducting gap energy.  In this work we investigate the low frequency conductivity in disordered superconducting NbN thin films close to the localization transition with time-domain terahertz spectroscopy.  In the normal state, strong deviations from the Drude form due to incipient localization are found.   In the superconducting state we find substantial spectral weight at frequencies well below the superconducting gap scale derived from tunneling.  We analyze this spectral weight in the context of a model of disorder induced broadening of the quasiparticle density of states and effective pair-breaking.  We find that although aspects of the optical and tunneling data can be consistently modeled in terms of this effect of mesoscopic disorder, the optical conductivity returns to the normal state value much faster than any model predicts.   This points to the non-trivial interplay of superconductivity and disorder close to localization.
  
 \end{abstract}

\maketitle

The manifestation of superconductivity in systems close to a disorder-driven localization transition has been an area of investigation for many years, yet many even central topics are not understood.  The electrodynamic response of such systems is a fundamental probe of their low-energy physics but wide-open issues exist here as well.  The optical conductivity corresponding to the mean-field Bardeen-Cooper-Schrieffer (BCS) theory of superconductivity was worked out in the context of the celebrated Matthis-Bardeen (MB) theory \cite{Matthisbardeen}.  A central prediction of the MB theory is the presence of a zero-frequency delta function and a gap 2$\Delta$ of a form that depends non-trivially on the BCS coherence factors in the real part of optical conductivity ($\sigma_1$). This theory works exceptionally well for many superconductors in the ``dirty" limit, where the normal state scattering rate ($1/\tau$) is much larger than the gap, but which are still far from a localization transition \cite{Tinkham56a,Palmer68a,Nuss91a}.  The MB theory predicts that in the limit of zero temperature, e.g. the gap is clean, there is no spectral weight in $\sigma_1$ for frequencies below 2$\Delta$. However, it has been noticed for many years that in highly disordered superconductors, for instance in thin-film systems near the superconductor-insulator transition, anomalous spectral weight develops near and below the expected gap edge. This has been observed in many different systems including granular superconductors \cite{Carr83a,Karecki83a,Stocker85a,Bentum86a,Pracht15a}, amorphous thin films \cite{Crane07a,Crane07b,Driessen12a}, and high-temperature superconductors with intrinsic disorder \cite{Corson00a,Orenstein03a}.

In this work we studied the low-frequency conductivity of strongly disordered superconducting NbN thin films close to the localization transition. In the normal state, strong deviations from the Drude form are found, which are indicative of incipient localization in these films. For medium disorder, the optical conductivity of the superconducting state is well-described by the MB formula. However, for higher disorder samples, additional low-energy spectral weight forms in a region below that predicted by the BCS theory.  For these samples, this energy is well below the scale of the gap determined by tunneling.  We investigated this feature in the context of prevailing models and conclude that its onset is reasonably described by a model of pair breaking from mesoscopic disorder.   However, discrepancies exist with the predicted shape of the conductivity in that in the most disordered samples, the conductivity recovers more quickly to the normal state values than predicted.   As the particular form of the MB conductivity functional derives from a particular form of the BCS coherence factors, this difference may presage a transition to an insulating state with localized Cooper pairs.

The low-frequency conductivity was measured with time-domain terahertz spectroscopy (TDTS) (See supplementary materials (SI)). The NbN used in this study consist of 60 nm and 120 nm epitaxial thin films that were grown by using pulsed laser deposition on (100)-oriented MgO single crystalline substrates.  Disorder in the NbN system can be tuned by varying the number of Nb vacancies in the crystalline NbN lattice \cite{Raychaudhuri11}.  Disorder introduced in these samples shows a homogeneous distribution at the nanoscale. The effective disorder in each film was quantified by the normal state conductivity just above the transition temperature (T$_c$) and calibrated to previous results \cite{Raychaudhuri11} that determined the room temperature $k_F l $, the product of the Fermi wave vector ($k_F$) and electronic mean free path ($l$).  Fig. 1 gives the details of T$_ c$ vs $k_F l $ for the samples used in this study.  At optimal deposition conditions NbN films have a T$_ c \approx$ 16 K. In our work we examined a range of films with $k_F l \sim   1.7 - 10.5$.  As disorder is increased T$_c$ decreases monotonically down to the limit where it is destroyed at $k_F l $ of order unity.

\begin{figure}[tb]
\scalebox{0.28} {\includegraphics [bb=330 30 19cm
19.5cm]{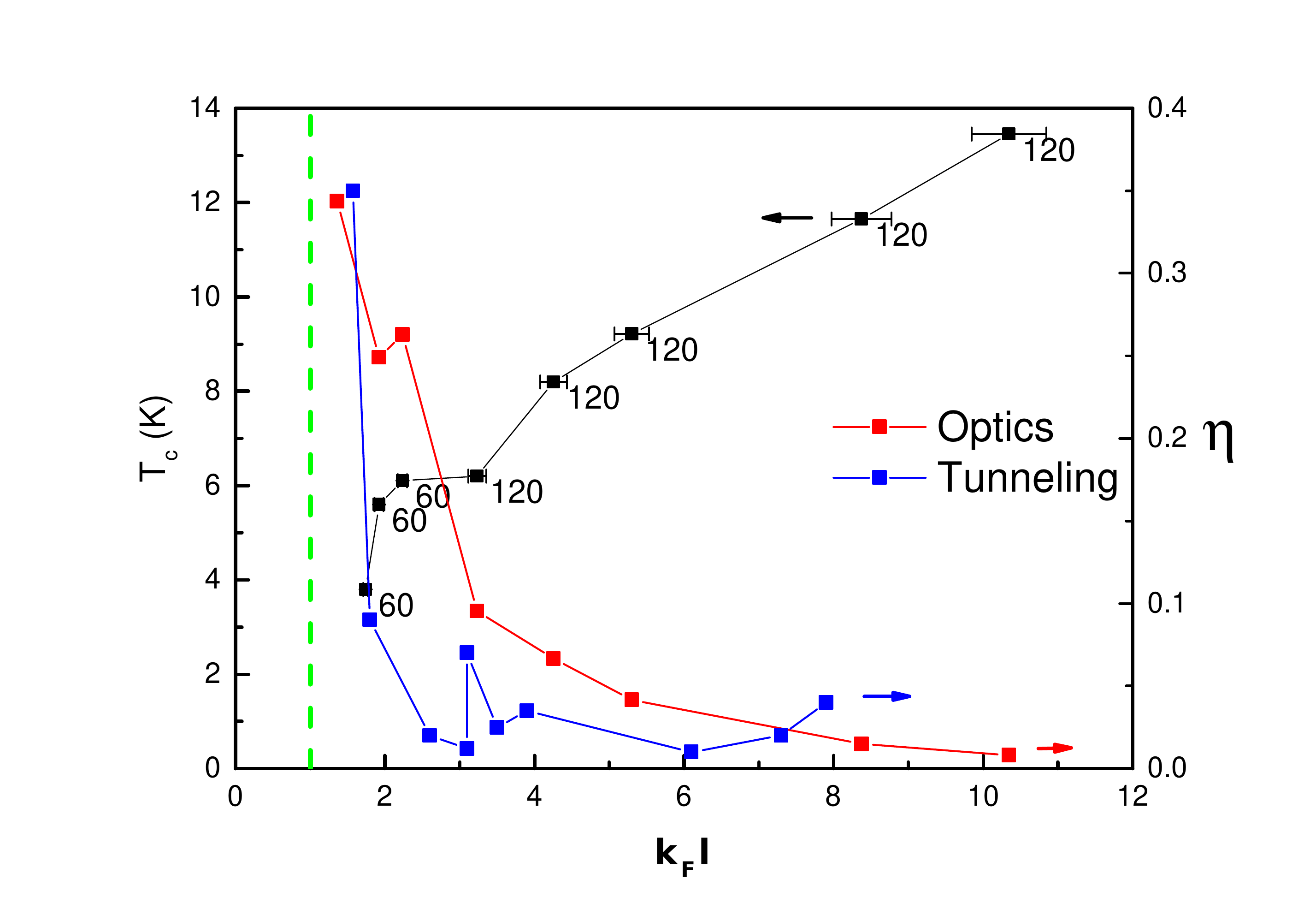}}
\caption{(Color online)   The left axis is T$_c$ vs. the dimensionless conductance parameter $k_{F}l$ for the samples used in this study. The thickness (unit is nm)of each sample is shown next to the data points.    T$_c$ was defined by the temperature where the resistance is indistinguishable from zero.  The green dashed line is $k_{F}l$ = 1.    On the right axis is $\eta$ extracted from optics and tunneling as discussed in the text.}
\end{figure}

In Fig. 2(a), we show the real parts of optical conductivity $\sigma_1$ just above T$_c$ ($\sim 1.1$ T$_c$) for this series of samples measured with TDTS. The spectra of the least disordered T$_c$ = 13.4 K sample is flat and featureless, indicating a Drude scattering rate that is much larger than the measured spectral range.  The spectra is consistent with dc transport measurements and indicates typical behavior \cite{Raychaudhuri09a} for a moderately disordered metal.  For increasing disorder, the normal state real conductivity is progressively  suppressed.  Even more significant for our analysis below is the deviations from conventional Drude behavior for samples with $k_F l  \lesssim   6$.  At higher disorder, one observes that the conductivity becomes a strongly increasing function of $\omega$, which is a signature of incipient localization in a disordered metal \cite{cooke06,Heinz11,Helgren02a}.  Consistent with this, dc transport has shown that as T$_c$ is suppressed, the resistivity of all lower T$_c$ samples show a negative temperature coefficient at low T \cite{Raychaudhuri09a}.  Localization modified Drude and Drude-Simith models have been proposed to include localization effects in such disordered systems and can reproduce the spectra with positive slope \cite{Heinz11}.

\begin{table*}
\caption{\label{tab:table2} Optical energy gaps, $2E_g$, are extracted directly from conductivity in Fig. 2 for each sample.  Units of $2E_g$ are in THz. $2E_g $/k$_B$T$_c$ is the ratios between the optical gap and the superconducting transition temperature. }
\begin{ruledtabular}
\begin{tabular}{ccccccccccccccccc}
 T$_{c}$&13.4 K&11.6 K&9.2 K
 &8.2 K&6.2 K&6.1 K&5.6 K&3.8 K\\
\hline
\textit{k}$_Fl$&10.35&8.37&5.30&4.25&3.23&2.24&1.92&1.74\\
$2E_g$(THz)& 1.10 & 0.92 & 0.66 
& 0.55 & 0.38 & 0.24&0.23& $<$0.12 \\

$2E_g $/k$_B$T$_c$& 3.93 & 3.81 & 3.45 
& 3.19 & 2.94 & 1.89&1.96& $<$1.52 \\

\end{tabular}
\end{ruledtabular}

\end{table*}

The real and imaginary parts of the optical conductivity at our lowest temperature of 1.5 K are shown in Figs. 2(b) and 2(c) for this series of samples.   For the highest T$_c$, a notable gap forms in $\sigma_1$ at the lowest temperature and a $1/\omega$ dependence is exhibited in $\sigma_2$.   (Please see the SI for all measured data for all measured samples).  As the disorder level increases, the optical gap decreases in accord with the lowering of T$_c$.   As the coefficient of the $1/\omega$ is set by the spectral weight in the zero frequency delta function, the coefficient of $\sigma_2$ decreases in accord with the delta function's dependence on both the gap and the normal state conductivity \cite{Matthisbardeen}.  It is important to note that despite the strong frequency dependence of the normal state conductivity due to localizing tendencies, the missing area that results from the formation of the gap, reappears in the spectral weight of the zero-frequency delta function.   However due to the strong frequency dependence of the normal state, this can only be seen by directly integrating the spectra and comparing the missing area to the coefficient of the $1/\omega$ part of $\sigma_2$ at low $\omega$.  It is also interesting to note that for the highest level of disorder, the high frequency parts of $\sigma_2$ show a progressively larger negative contribution.   This negative contribution is also apparent in the normal state and comes from the increasing relative effect of finite frequency excitations on the real part of the low-frequency dielectric function (e.g. the system polarizability) and departures from the Drude form due to the localizing tendencies of the normal electrons.


\begin{figure}[t]
\scalebox{0.45} {\includegraphics [bb=330 30 11cm
17cm]{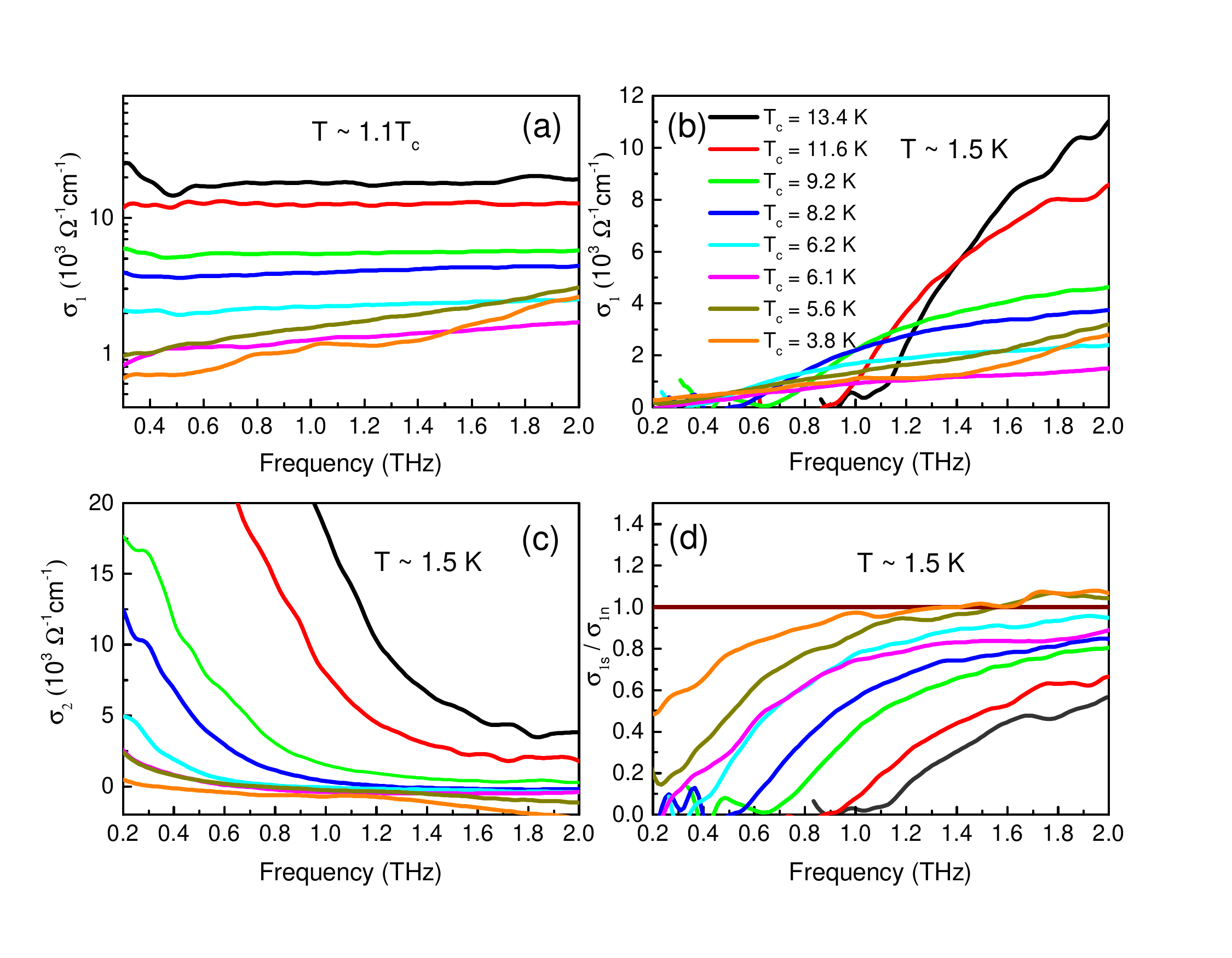}}
\caption{(Color online)  (a) Real part of the optical conductivity at 1.1 T$_{c}$. (b) Real part and (c) Imaginary parts of the optical conductivity at 1.5 K. (d) Real parts of optical conductivity at 1.5 K normalized by the normal state conductivity at 1.1 T$_{c}$ given in (a).}
\end{figure}

\begin{figure}[t]
\scalebox{0.45} {\includegraphics [bb=330 30 11cm
17cm]{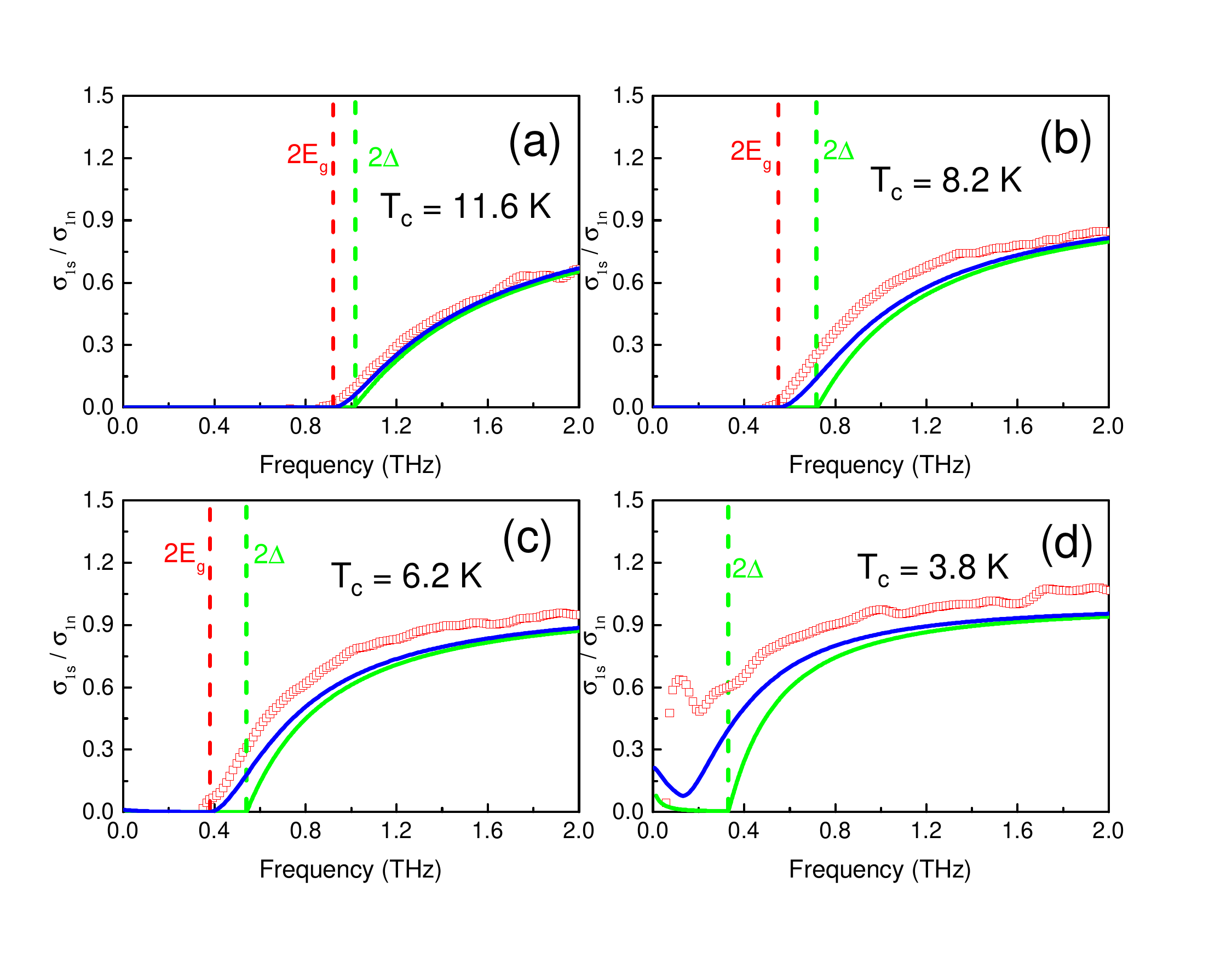}}
\caption{(Color online)  The red hollow squares represent the real parts of the optical conductivity at 1.5 K for four representative samples.  The red dashed vertical curves show the optical energy gaps directly extracted from optics. The green dashed lines indicate the superconducting gaps extracted from tunneling.  The green curves are created via a numerical solution to the MB formalism with superconducting gaps extracted from tunneling. The blue curves are simulations using the model of Larkin and Ovchinnikov. }
\end{figure}

In the conventional MB theory, the real part of the superconducting state conductivity is expressed as a ratio to the normal state conductivity to normalize out the transition matrix elements between the single electron states. However, in the usual theory it is expected that this normal state real conductivity is flat in frequency and its imaginary part is zero as is typical for a highly disordered metal with a large scattering rate.  Due to the strong deviations from the expectation in the normal state conductivity we found that (except for the least disordered sample), it is completely impossible to simultaneously fit both complex components to the MB form when using a frequency independent $\sigma_n$ even when letting the gap be a free parameter.

To eliminate the localizing features in the spectra, we normalized the real part of the superconducting state conductivity by the normal state real conductivity (1.1 T$_c$) (Fig. 2(d)). The optical energy gap $2E_g$ can be extracted from these normalized conductivity directly as the minimum or threshold in $\sigma_1$.  Here and in what follows, we use $2E_g$ to differentiate the optical gap from the gap measured in tunneling.  Traditional BCS theory predicts that the ratio between the optical gap and transition temperature should be 3.5, while strong coupling effects can drive it larger.  As shown in Table 1, for our lowest disorder T$_c$ = 13.4 K sample, the ratio between optical gap and transition temperature 2$E_g $/k$_B$T$_c$ is 3.93. As T$_c$ is suppressed to 8.2 K, the ratio falls below the BCS stability limit of 3.53. For the T$_c$ = 3.8 K sample, a clear minimum or threshold cannot be seen in the conductivity $\sigma_1$ in superconducting state (Fig. 1(8a) of SI).  Considering the low detection limit of our spectrometer ($\approx$ 0.12 THz), we estimate 2$E_g $/k$_B$T$_c < 1.5$ for this sample. It is interesting to compare these numbers to those extracted from tunneling.  Tunneling spectra in moderately disordered conventional superconductors reveals a conventional BCS density of states with its square root singularities and a clean gap.   With increasing disorder -- reminiscent of the situation in optics -- the density of states broadens \cite{Dynes84,Raychaudhuri13,Sacepe11a} and although the energy separating the coherence peaks ($2\Delta$) maintains a ratio $2\Delta$/k$_B$T$_c \approx 4$ up to high disorder levels \cite{Raychaudhuri09a}, the peaks become smeared and a tunneling conductance develops at lower energies (See SI).

In Fig. 3, we show the normalized optical conductivity for four typical samples. The red dashed lines label the positions of optical energy gaps extracted by inspection.  They can be compared to the green dashed lines that indicate the expected superconducting gap 2$\Delta$ which were given (as it is in the MB theory) by the experimentally determined relation 2$\Delta$/k$_B$T$_c = 4.2$ with 2$\Delta$ the energy gap determined from fits to tunneling (See SI).   Using the MB theory, we simulate the normalized conductivity (green curve in Fig. 3) with these superconducting gaps from tunneling.  With increasing disorder, additional spectral weight progressively develops both below and above the gap scale 2$\Delta$.  For T$_c$ = 3.8 K, a clear energy gap could not be observed from the conductivity spectra.  It is also observed that for the most disordered samples the normalized conductivity approaches the normal state faster than the BCS prediction. 

A number of possibilities beyond MB theory exist to explain this anomalous absorption.  In principle, both collective modes and quasiparticle excitations may contribute. It has been pointed out that, in systems with a spatial modulation of the superfluid density, one may find absorptions far below 2$\Delta$ \cite{vanderMarel96a,Barabash03a,Orenstein03a}.  Cea et al. give a similar scenario where the spectral weight may be exhibited at finite frequency \cite{Cea14a}.  In all cases these are low-energy phase-like modes that are rendered optically active at $q\sim 0$ by the breaking of translational symmetry in the strongly disordered system.  In the present case, we do not believe phase modes are the obvious choice to explain most of the additional absorption because it is appears to be associated with gap edge excitations, which is not necessarily a relevant energy for phase degrees of freedom.  We believe that if phase mode absorptions are significant, it is only on the most disordered samples (T$_c$ $<$ 3.8 K) do not have clean gap within our measured $\omega$ range.

Alternatively, it was recently claimed in the study of the optical response of disordered thin films that the in-gap optical conductivity exhibited a sharp threshold that was consistent with an excitation of the amplitude of the order parameter \cite{Sherman15a}.  This amplitude mode, if it exists, is an analog to the famous Higgs boson from particle physics.  However, amplitude modes as such are not generically guaranteed in condensates \cite{Pekker15a}, and in a BCS-style superconductor, amplitude modes are over damped as they are degenerate with the quasiparticle absorption edge at $2 \Delta$.  The interpretation in Ref. \onlinecite{Sherman15a} was made on the basis of a specific particle-hole symmetric $O(2)$ relativistic field theory\cite{Podolsky11a} where the quasiparticle energy scale is set to infinity.  It is not clear how the physics of this $O(2)$ field theory connects to the BCS limit, which is obvious in our data for $k_F l \gg 1$.  Moreover, in all known circumstances in which the amplitude mode threshold can be pushed below the quasiparticle absorption edge and rendered optically active, e.g. in the limit of strong disorder or strong-coupling, particle-hole symmetry is broken which forces amplitude and phase modes to mix and a clean distinction between the excitations in different sectors is obviated.  As pointed out in Ref.  \onlinecite{Cea15a} there are even internal consistency issues with the possibility to see an amplitude mode optically.  Because the scalar amplitude mode only becomes optically active by being excited in conjunction with a phase mode, a coupling between sectors is necessary for an amplitude mode's observation - yet this very coupling renders the amplitude and phase modes indistinct.  Note that none of our data shows either the sharp onset or the particularly low-energy scale of the single displayed high disorder curve in Ref. \onlinecite{Sherman15a}.  

Irrespective of the above considerations, it is clear that, with substantial tunneling conductance below $\Delta$, it is inadequate to model the optics with an MB functional that relies on a clean gap.  Tunneling measurements are an important point of comparison to optical conductivity by virtue of the fact that they probe quasiparticle effects directly and only indirectly probe collective modes through their coupling to quasiparticles.  We propose that the low threshold (as compared to T$_c$) of 2$E_g$ that we see in optics derives from the same sub-gap states seen below the coherence peak $\Delta$ in tunneling.  It is quite natural to expect a modification of the quasiparticle excitations of the system at high disorder.  Larkin and Ovchinnikov showed that that disorder in the form of a spatially varying BCS coupling constant will give an effective pair breaking effect \cite{AO} that maps to the Abrikosov-Gor'kov pair breaking model caused by magnetic impurities \cite{AG}.  A similar mechanism may be applicable to superconductors with mesoscopic fluctuations \cite{skvortsov12}.

We can model both quasiparticle properties of the optical and tunneling data in a self-consistent fashion by the model of Larkin and Ovchinnikov (LO) \cite{AG,AO}.  The prediction was that the density of states would be homogeneously broadened from the BCS expectation with an energy gap  renormalized to $E_g (\eta)=(1-\eta^{2/3})^{3/2} \Delta$. Here $\Delta$ is the average value of order parameter (very approximately indicated in the tunneling by the energy of the coherence peaks) and $\eta$ is a parameter that sets the strength of the effective depairing \cite{Comment2}. By using $\Delta$ extracted from tunneling and $E_g$ from optics, we estimate $\eta$ for each sample we studied and plot them on the right side of Fig. 1.  As $k_{F}l$ decreases, $\eta$ increases.   Although this method can qualitatively explain the lower threshold, the values of $\eta$ are systematically larger than what is predicted from theory at these  $k_{F}l$ values \cite{skvortsov12}.  In this regard, the mesoscopic fluctuations may be regarded as the minimal model of disorder and other types of  microscopic inhomogeneity may push $\eta$ higher.  Irrespective of this, we can compare these $\eta$'s with those extracted from direct fits of the LO model to the tunneling conductance \cite{Chand12a}.   One can see that although the values of $\eta$ extracted by the two methods are close, optics gives a value systematically higher.  This is consistent with both recent experiments that compared the $\eta$ determined from the superfluid density with that of tunneling \cite{Driessen12a} and recent theory \cite{skvortsov16} that predicted (for the 2D case, which is not necessarily applicable in our thick films) that the $\eta$ from optics should be generally larger in this disorder range by factor of 6/ln(6$g^2$) (with $g$ the dimensionless conductance) due to the role of vertex corrections in transport.

To more precisely compare the LO model to the data we solved the Usadel equation $i E\mathrm{sin}\theta + \Delta \mathrm{cos}\theta - \eta \Delta \mathrm{sin}\theta \mathrm{cos}\theta = 0 $ numerically with $\Delta$ taken from tunneling and $\eta$ is estimated above.  Here, $E$ is the energy relative to Fermi level, $\theta$ is the pairing angle and sin$\theta$ and cos$\theta$ are the disorder-averaged Green's functions \cite{Driessen12a}. The single particle density of states is directly given by $\rho(E)= \rho_0$Re(cos$\theta$), where $\rho_0$ is the normal state density of states. We show our simulations of the density of states in the part D of the SI. The corresponding normalized real optical conductivity at T = 1.5 K can be calculated through the expression

\vspace{-4mm}

\begin{equation}
\begin{aligned}
\frac{\sigma_{1s}}{\sigma_{1n}}=&\frac{2}{\hbar \omega} \int_{E_{g} }^{\infty}[f(E)-f(E+\hbar \omega)]\mid F(E,E+\hbar \omega)\mid dE +   \\
&\frac{1}{\hbar \omega}\int_{E_{g} - \hbar\omega}^{-E_{g}} [1-2f(E+\hbar \omega)]\mid F(E,E+\hbar \omega)\mid dE
\end{aligned}
\nonumber
\end{equation}

\vspace{-1mm}

\noindent where the generalized coherence factor is given by $F(E,E + \hbar \omega)$ = Re[$\mathrm{cos} \theta(E)$] Re[$\mathrm{cos}\theta(E + \hbar \omega$)] + Im$[\mathrm{sin} \theta(E$)] Im$[\mathrm{sin}\theta(E + \hbar \omega$)].   Here $f(E)$ is Fermi-Dirac distribution function.  In the limit where $\eta$ = 0, one recovers the traditional MB form. We show the simulation of normalized conductivity in Fig. 3 (blue). As expected from the above, after considering broadening effects around the gap edge in the density of states, a notable amount of optical spectral weight fills the region between 2$\Delta$ and 2$E_g$.  At high frequency, simulation with the LO model recovers the predictions of MB.  Our simulation qualitatively explains the conflicts between optics and tunneling or rather demonstrates that when making a comparison one cannot compare the threshold in optics to the energy of the coherence peaks.  Although our model successfully accounts for the lower onset energy of the optical gap as compare to tunneling, the theoretical curves still do not capture the high-frequency parts of normalized conductivity.  We find that in the most disordered samples, the conductivity recovers more quickly to the normal state values than predicted.   As the particular form of the MB conductivity functional derives from a particular form of the BCS coherence factors, this difference may presage a transition to an insulating state with localized bosonic Cooper pairs. However, we cannot rule out that this feature does not come from our normalization procedure where we divide by the strongly frequency dependent conductivity. Although calculations have been done showing the role that mesoscopic disorder plays in suppressing the superfluid density \cite{skvortsov16}, no explicit calculation of the gap edge structure has been performed.   Moreover calculations of the gap-edge optical response across the BEC-BCS crossover have not been made.  Such contributions would be very welcome.

We would like to thank D. Arovas, A. Auerbach, L. Benfatto, M. Feigel'man, U. Pracht, M. Randeria, M. Scheffler, M. Skvortsov, N. Trivedi, and C. Varma for discussions. We thank Dr. Bing Xu for his kind help in coding. The research at JHU was supported by the NSF DMR-1508645 and the Gordon and Betty Moore Foundation through Grant No. GBMF2628 to NPA.   Research at Tata Institute of Fundamental Research was supported by Department of Atomic Energy, Government of India.


\newpage

\setcounter{figure}{0}
\setcounter{equation}{0}
\setcounter{section}{0}
\begin{widetext}

\textbf{Supplemental Material: Anomalous gap edge dissipation in disordered superconductors on the brink of
localization}

\bigskip

\section{A. Time-domain terahertz spectroscopy}

A home-built time-domain terahertz (THz) spectrometer was used for this work.  GaAs Auston switches are used as emitters and receivers to generate and detect THz pulses. An ultrafast laser (800 nm) is split into two paths by a beamsplitter. One beam travels to the biased emitter and generates a THz pulse. This THz pulse passes through the sample or substrate and arrives at the receiver. The other laser beam propagates to the receiver and is used to gate the THz pulse coming from the emitter side. The beam path difference between these two laser beams is precisely controlled by a delay stage to map out the E field as a function of time of the THz pulse. By mapping out THz pulses after transmitting through substrates and samples separately, and taking a ratio of the Fourier transforms, we obtain transmission function in the frequency domain. The complex conductivity of thin films can be directly calculated in the thin-film limit with the expression \cite{Luke,Valdes,Liang}

\begin{equation}
T(\omega) = \frac{1+n}{1+n+Z_{0}d\sigma(\omega)}e^{\frac{i\omega (n-1)\Delta L}{c}}.
\end{equation}

\noindent Here $T(\omega)$ is the transmission as referenced to MgO substrate, $\sigma$ is the complex optical conductivity,  $n$ is the refraction index of substrate, $\Delta L$ is the small thickness difference between samples and reference substrates, $d$ is the thickness of the thin film and $Z_{0}$ is the vacuum impedance.

\section{B. Temperature dependent conductivity of NbN thin films}

Temperature dependent real and imaginary parts of the optical conductivity of all measured NbN thin films are shown in Figure 1. For the T$_{c}$ = 13.45 K sample, the normal-state conductivity shows a standard metallic behavior. But as T$_{c}$ is suppressed to 8.2 K by disorder, the real parts of the optical conductivity in the normal state begin to display a positive slope with frequency. With T$_{c}$ further suppressed, this positive slope behavior becomes more notable. At the same time, the imaginary part of the optical conductivity gets a negative contribution. All these phenomena indicate that the localization of the normal state carriers in strongly disordered NbN thin films play an important role in transport.  Our study reflects the interesting interplay between an incipient localization and superconductivity.

\begin{figure}[H]
\includegraphics[width=0.5\columnwidth]{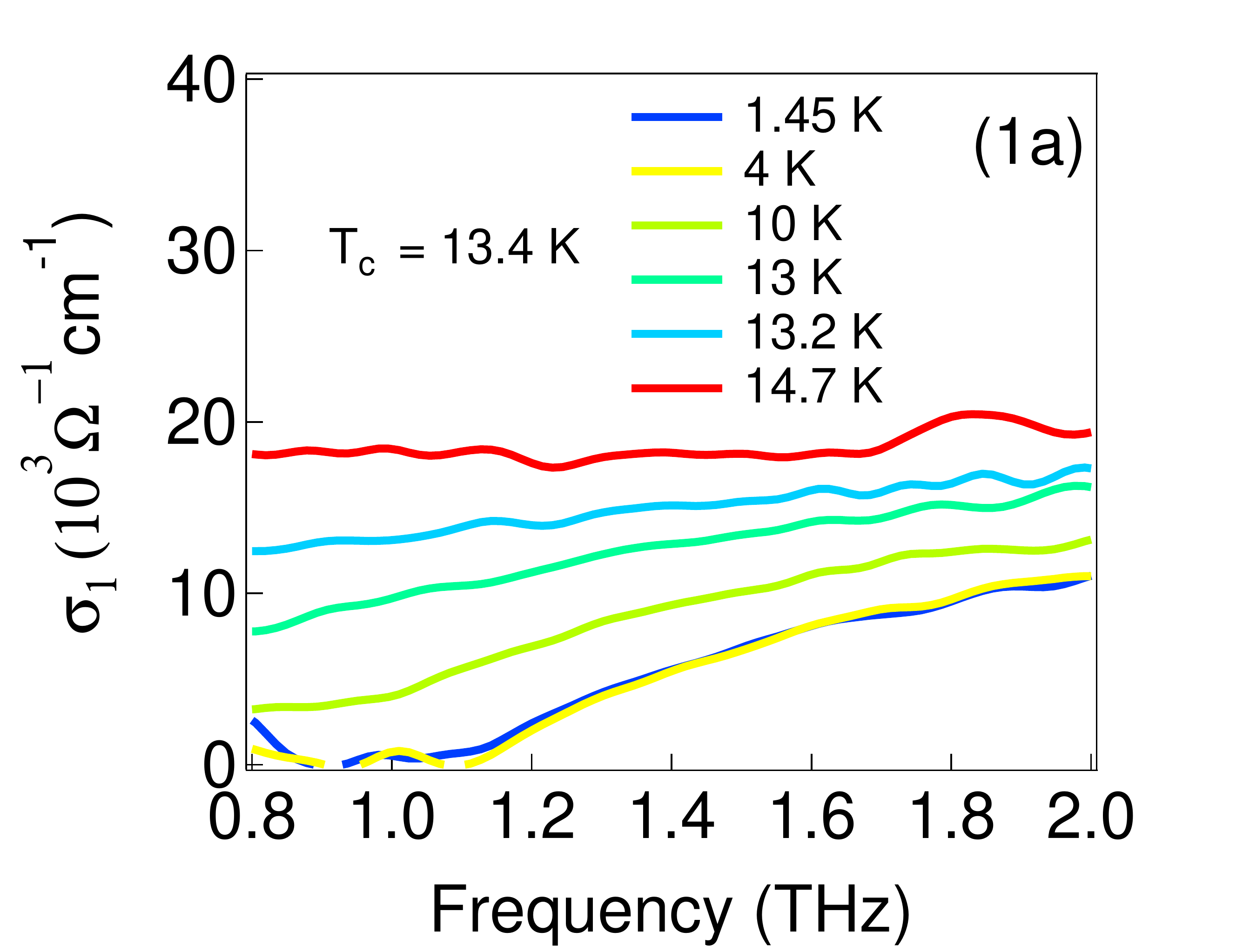}
\includegraphics[width=0.5\columnwidth]{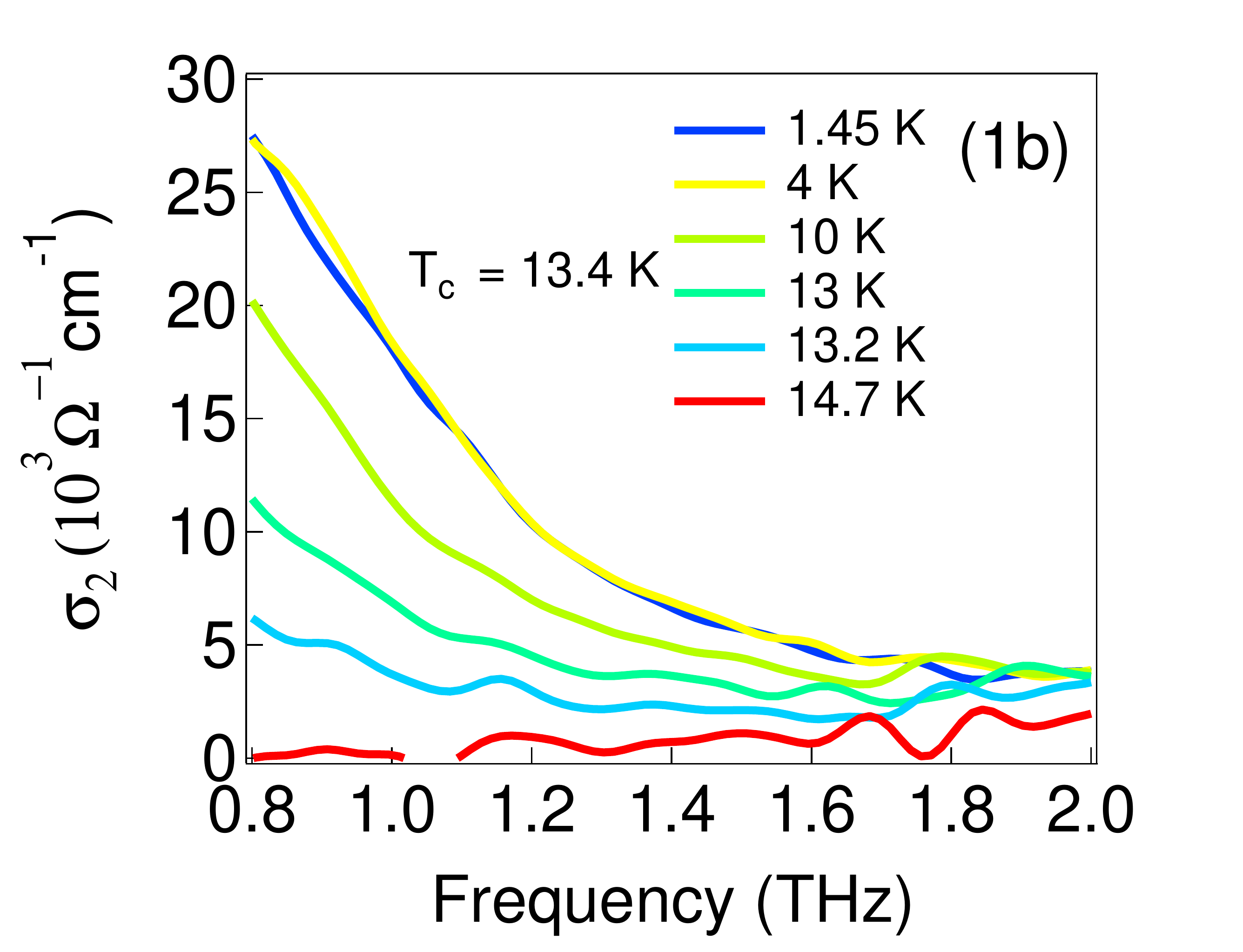}
\includegraphics[width=0.5\columnwidth]{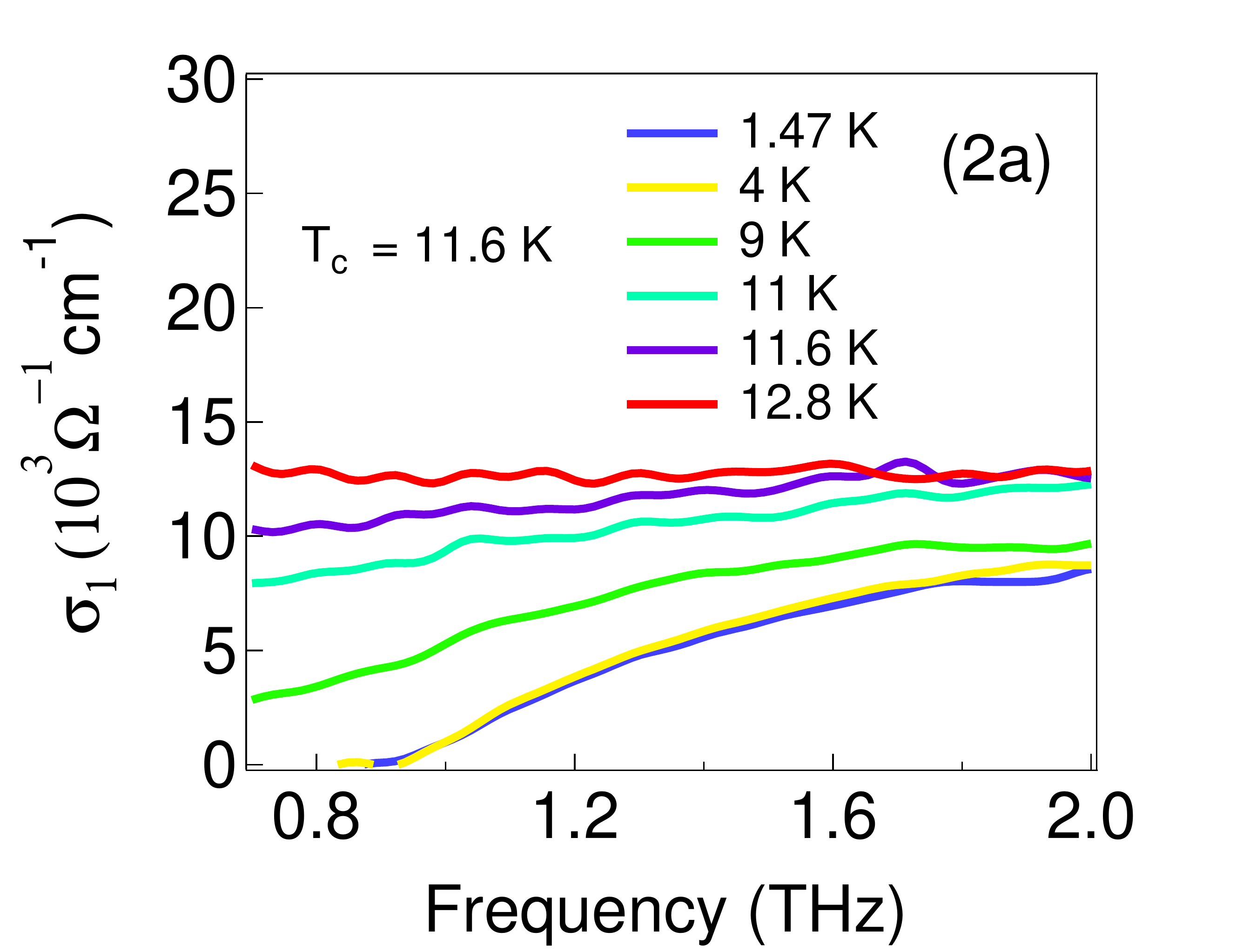}
\includegraphics[width=0.5\columnwidth]{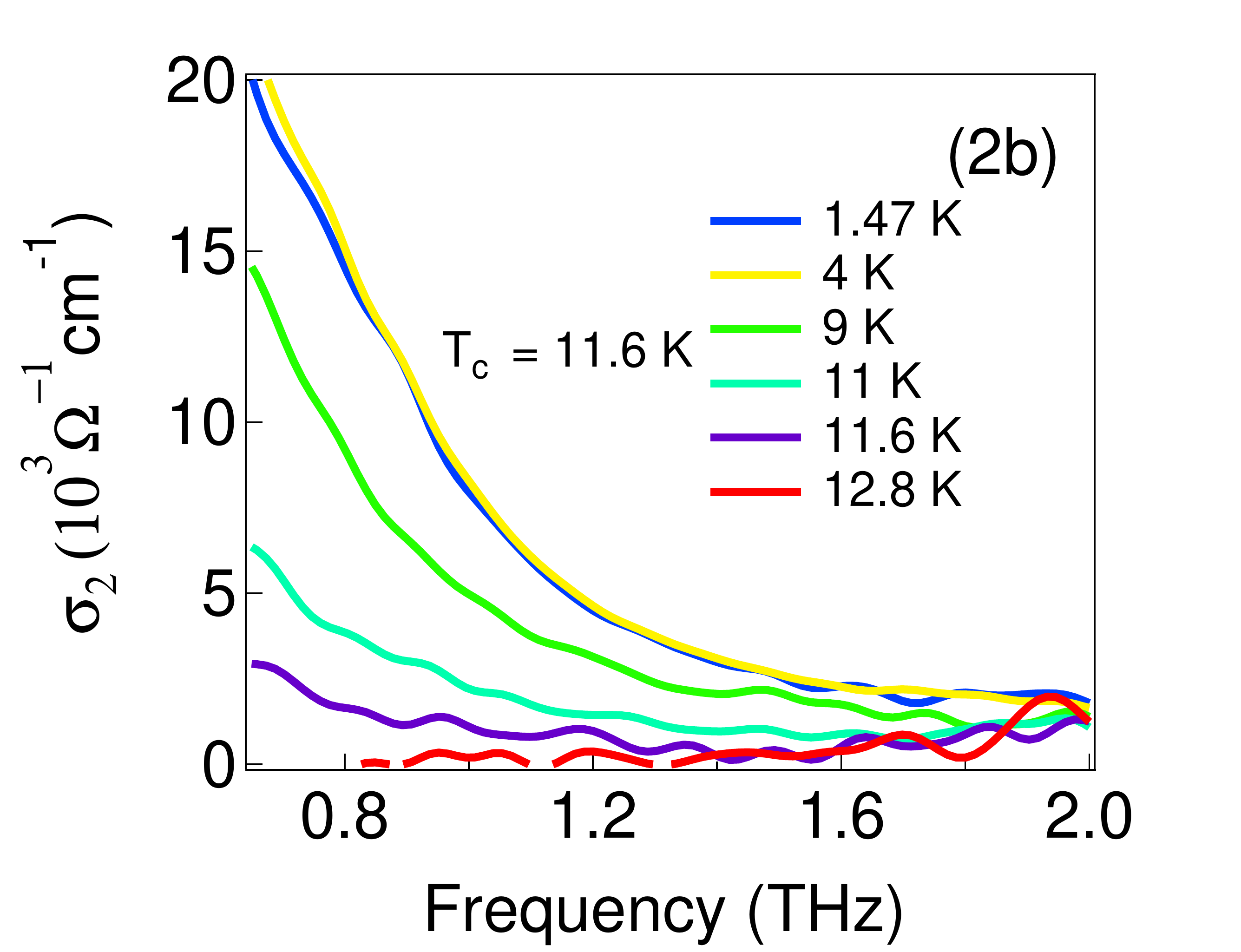}
\includegraphics[width=0.5\columnwidth]{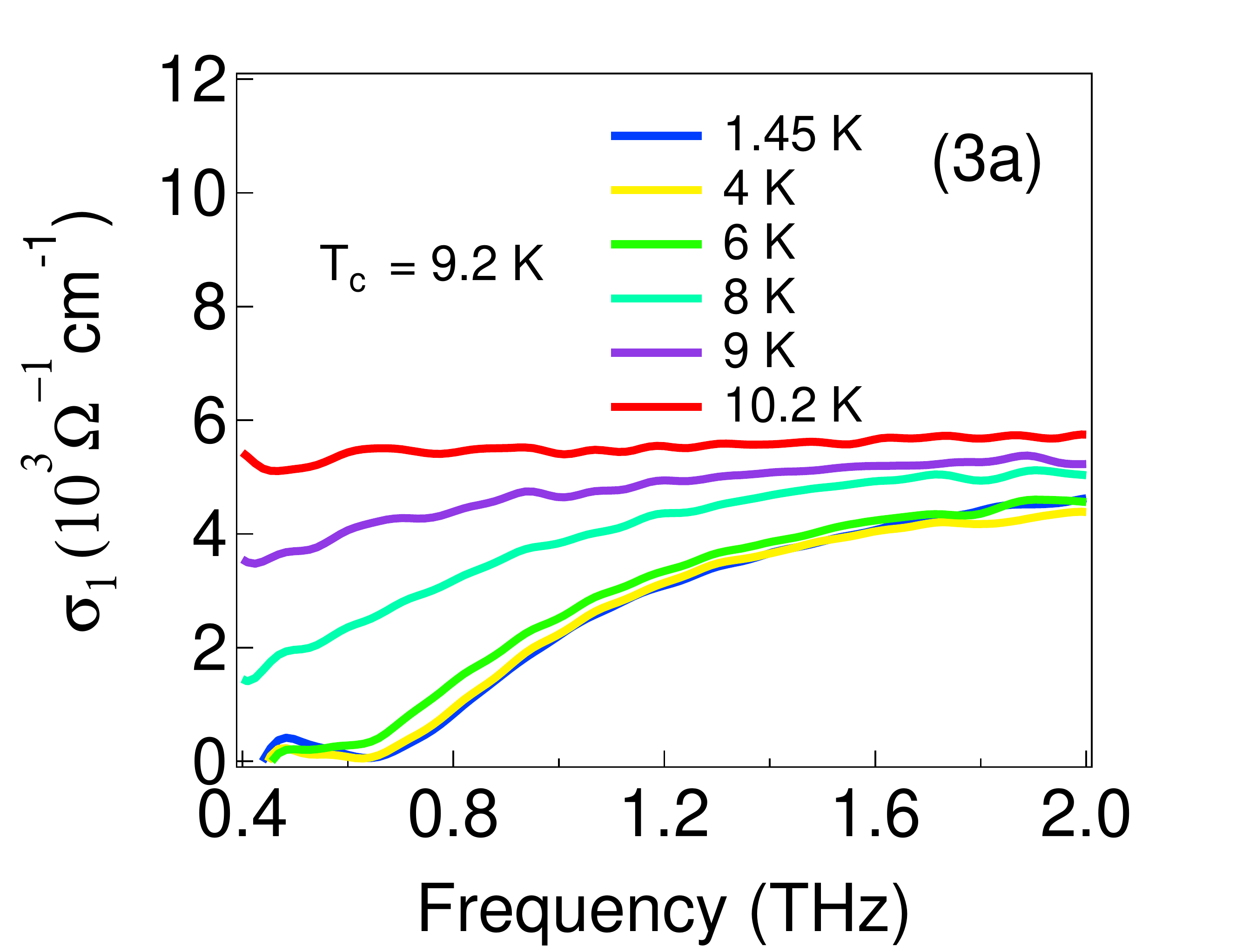}
\includegraphics[width=0.5\columnwidth]{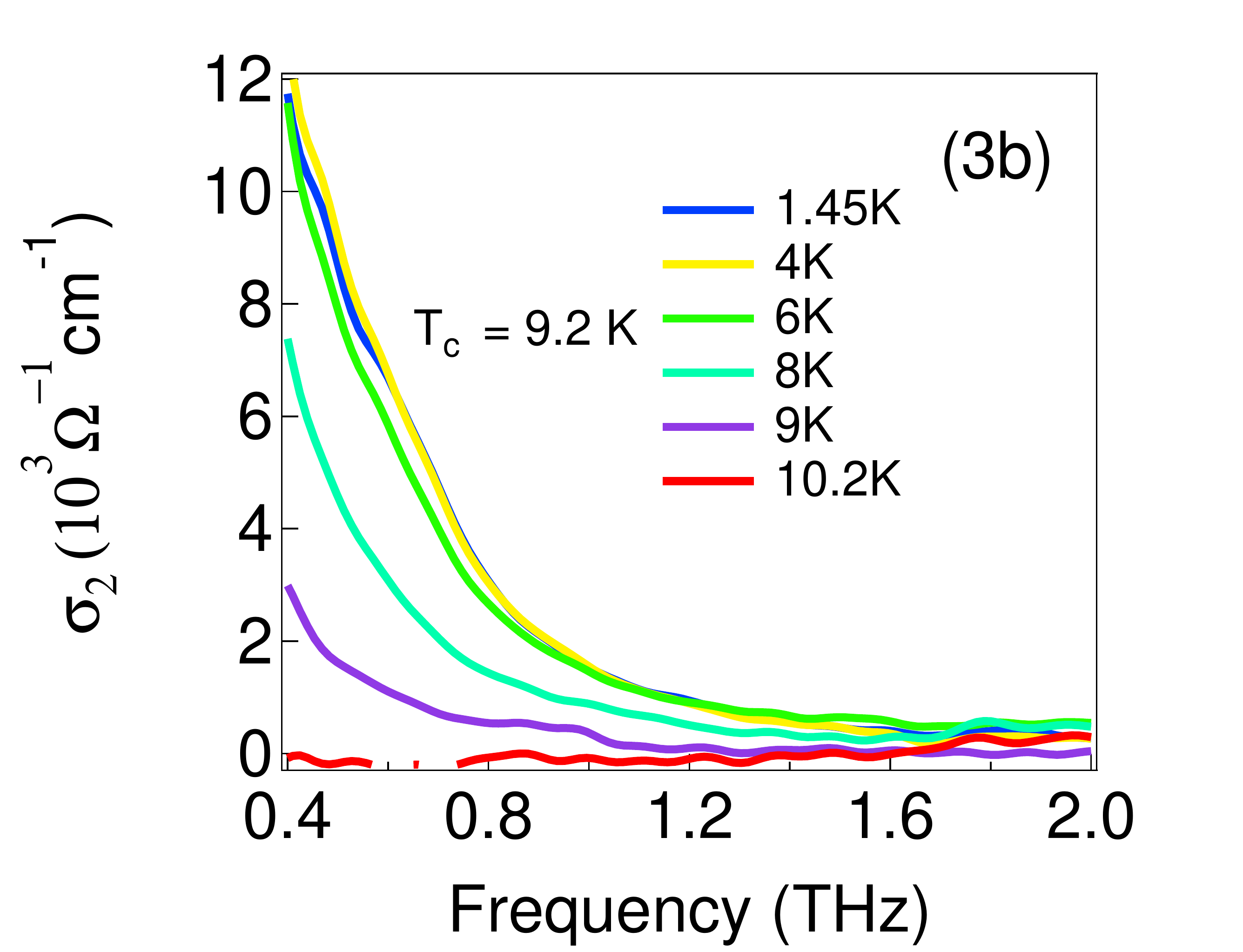}
\end{figure}

\begin{figure}[H]
\includegraphics[width=0.5\columnwidth]{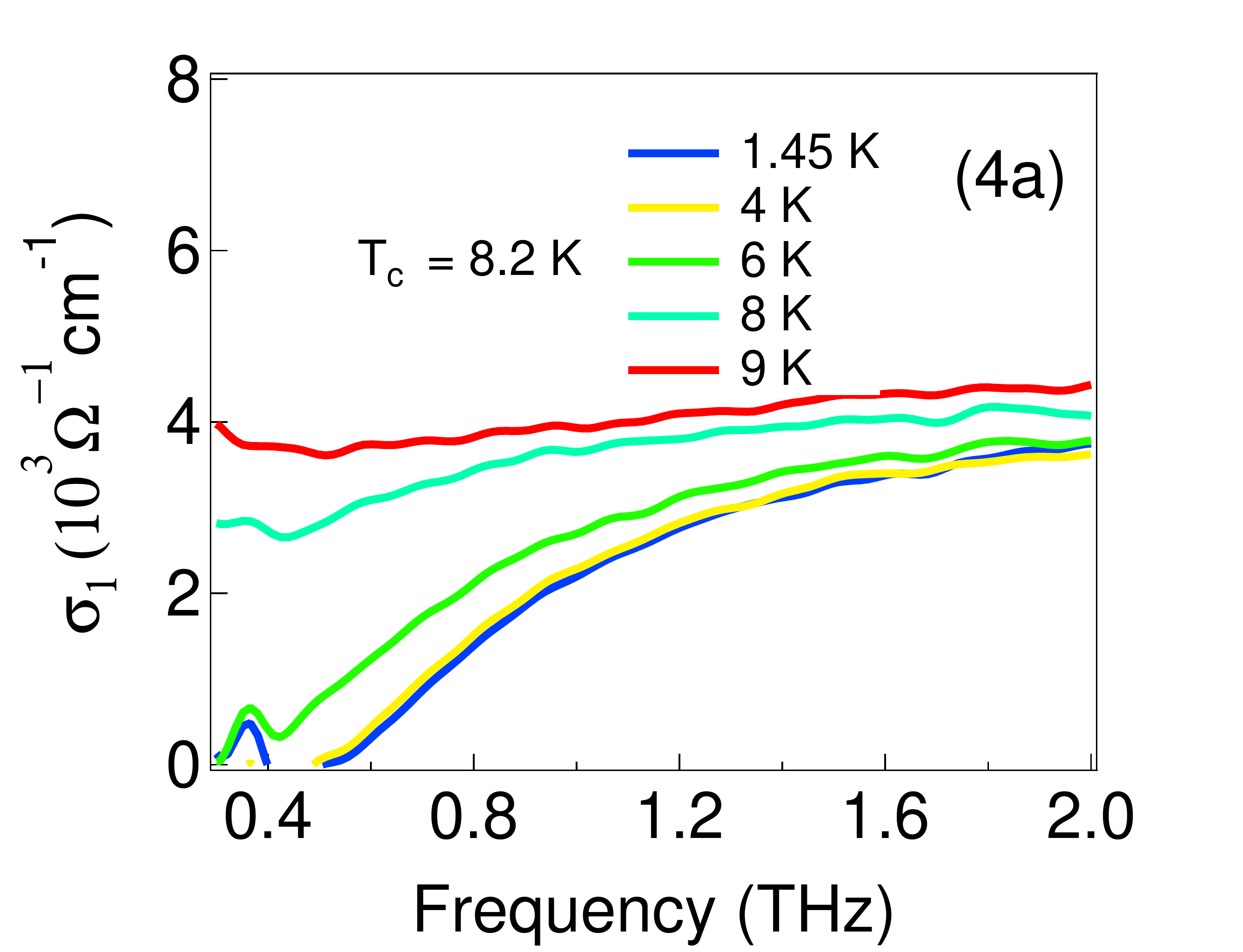}
\includegraphics[width=0.5\columnwidth]{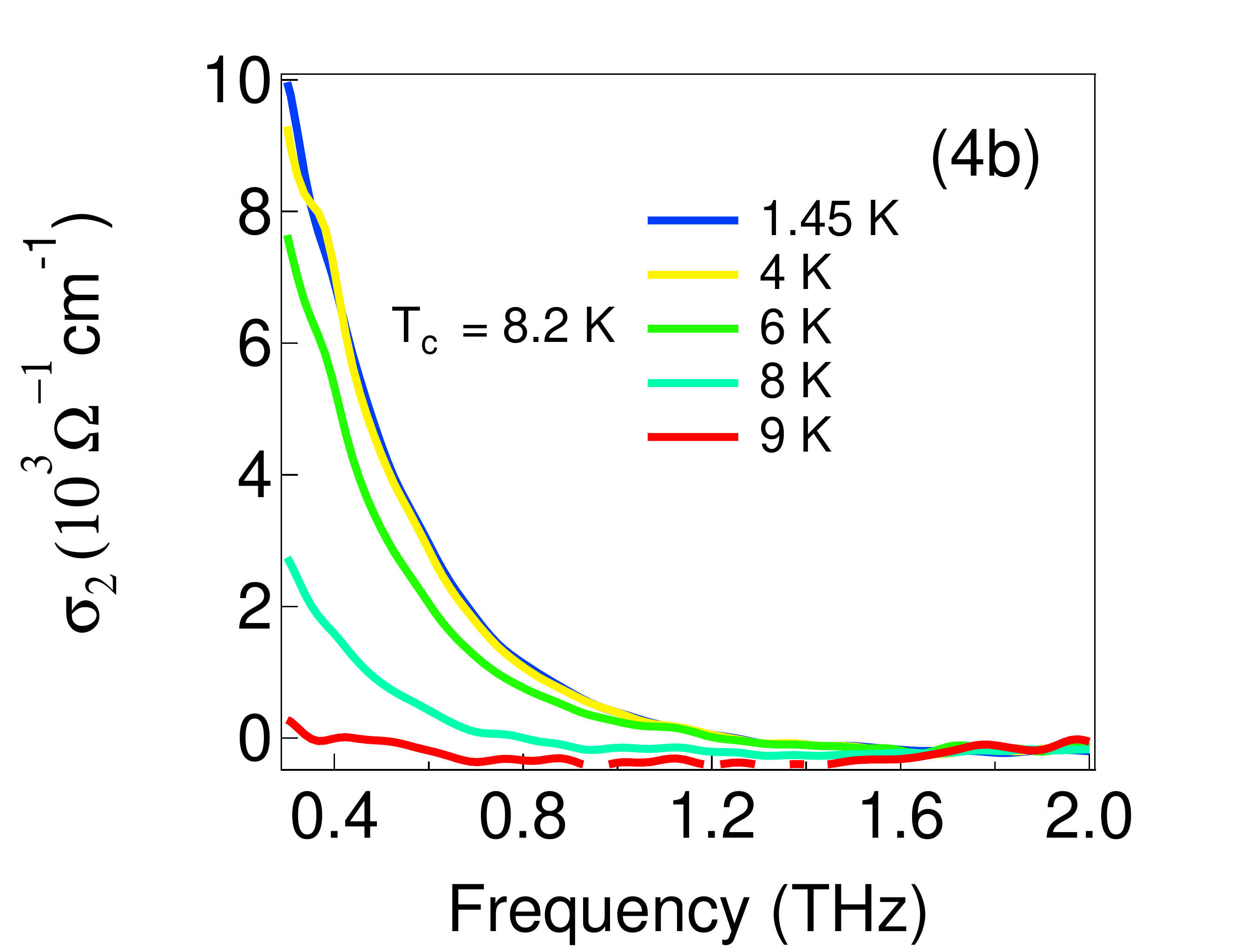}
\includegraphics[width=0.5\columnwidth]{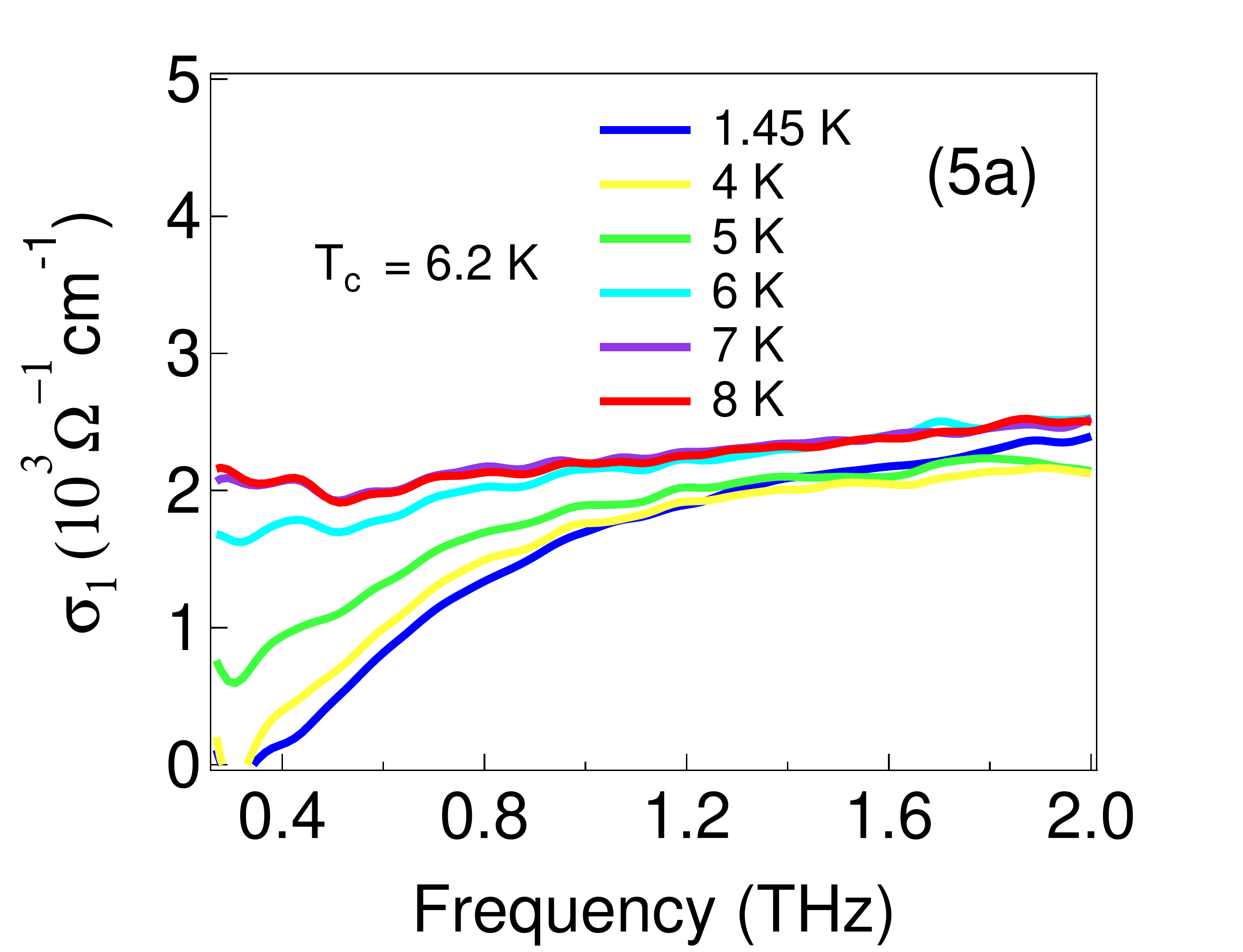}
\includegraphics[width=0.5\columnwidth]{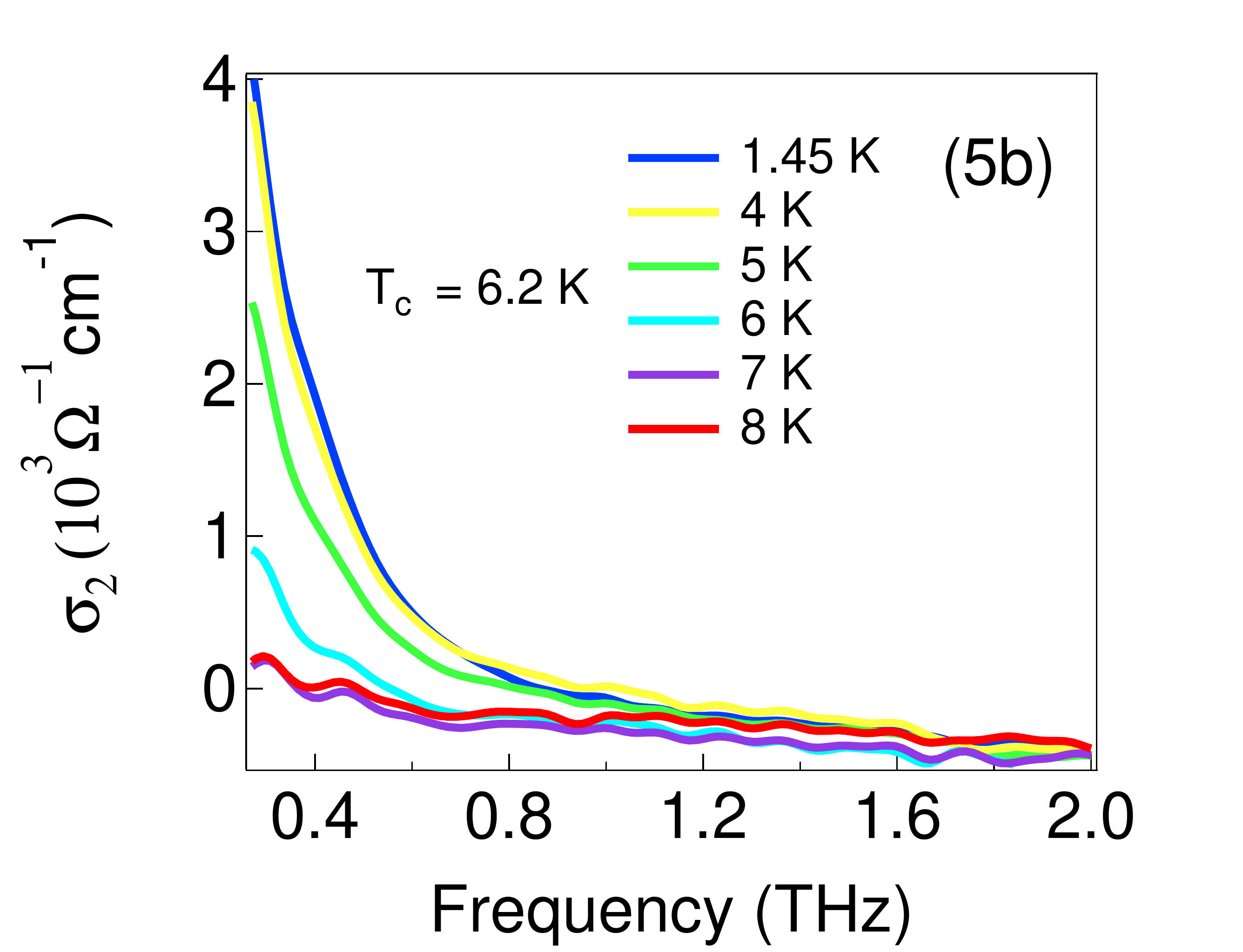}
\includegraphics[width=0.5\columnwidth]{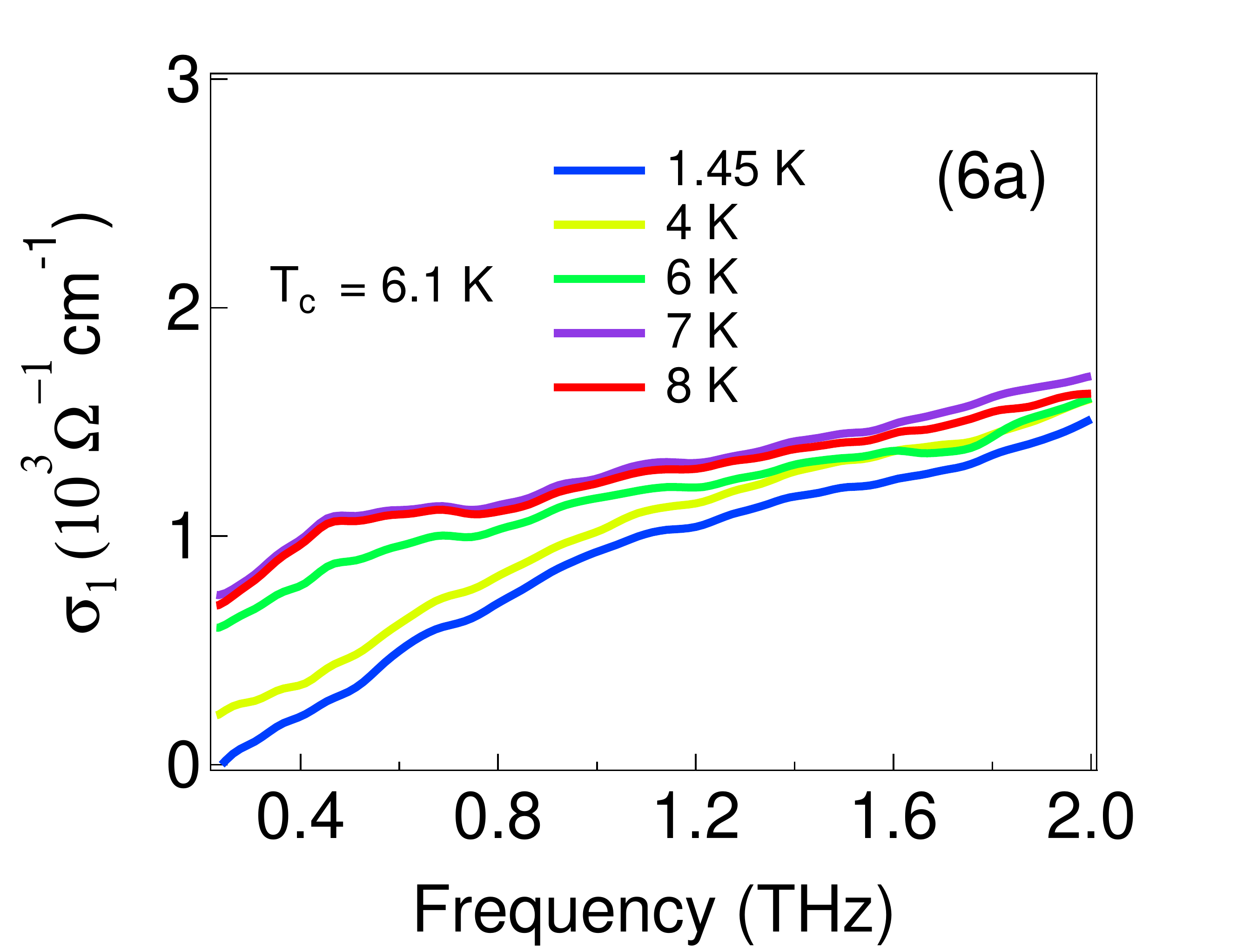}
\includegraphics[width=0.5\columnwidth]{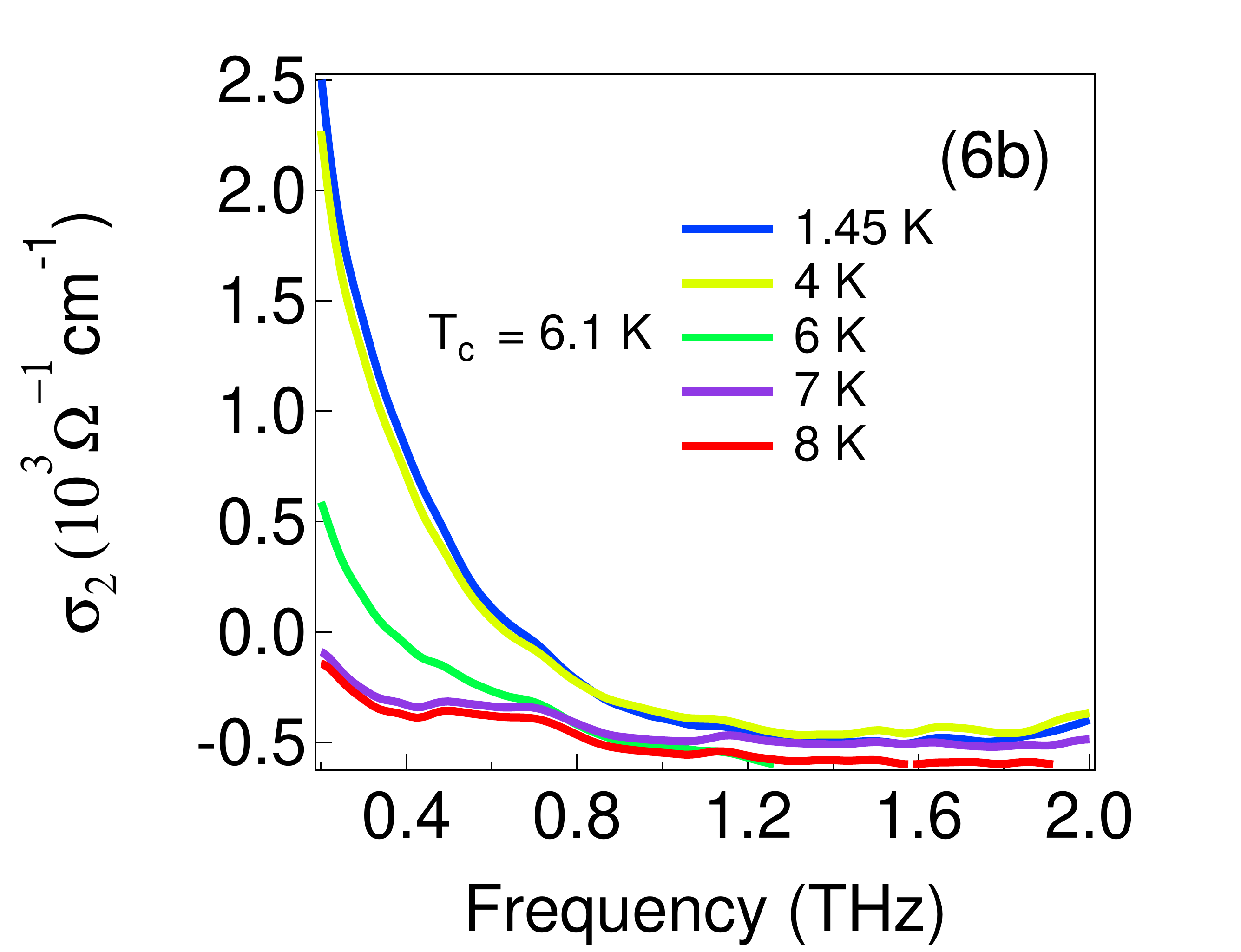}
\end{figure}

\begin{figure}[H]
\includegraphics[width=0.5\columnwidth]{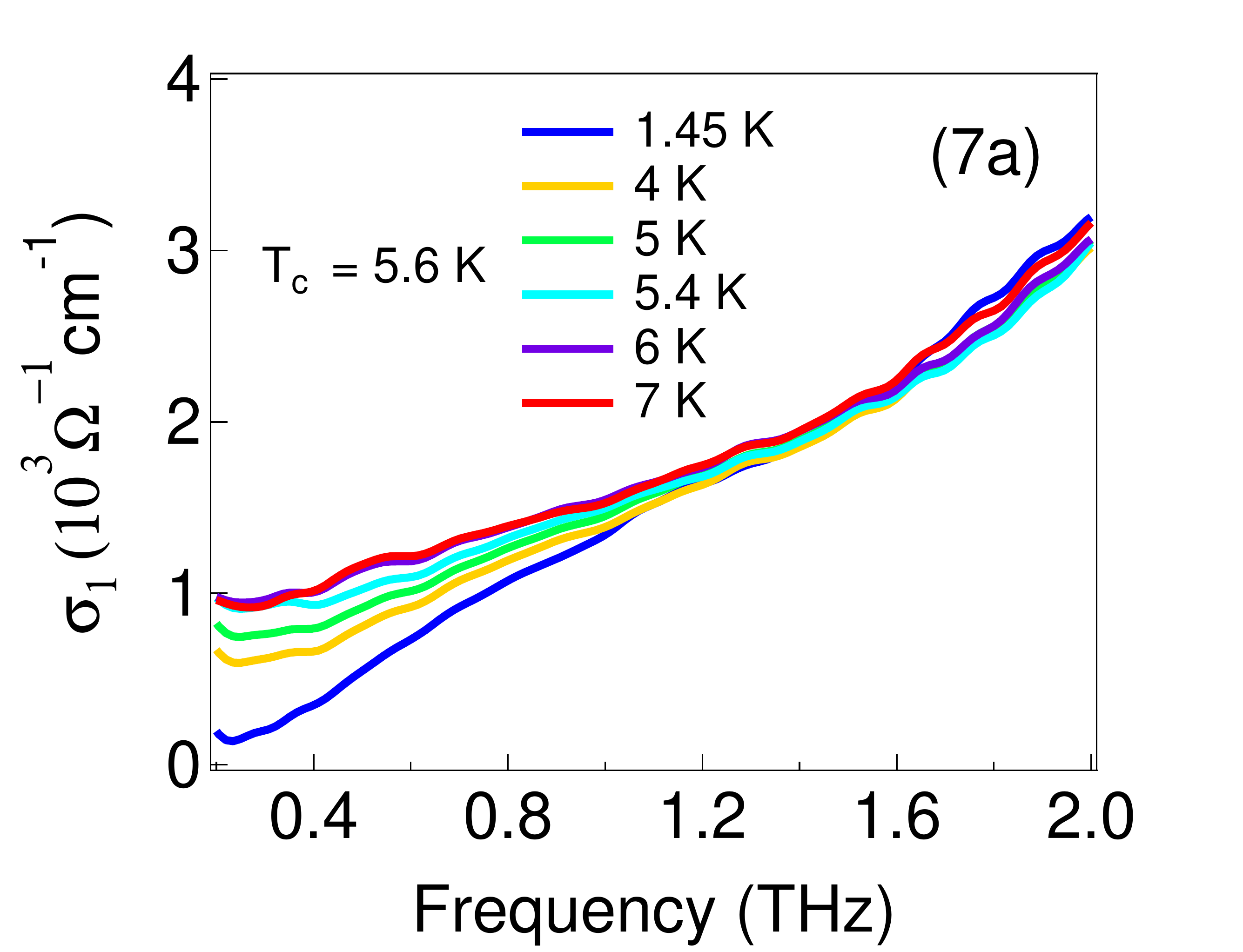}
\includegraphics[width=0.5\columnwidth]{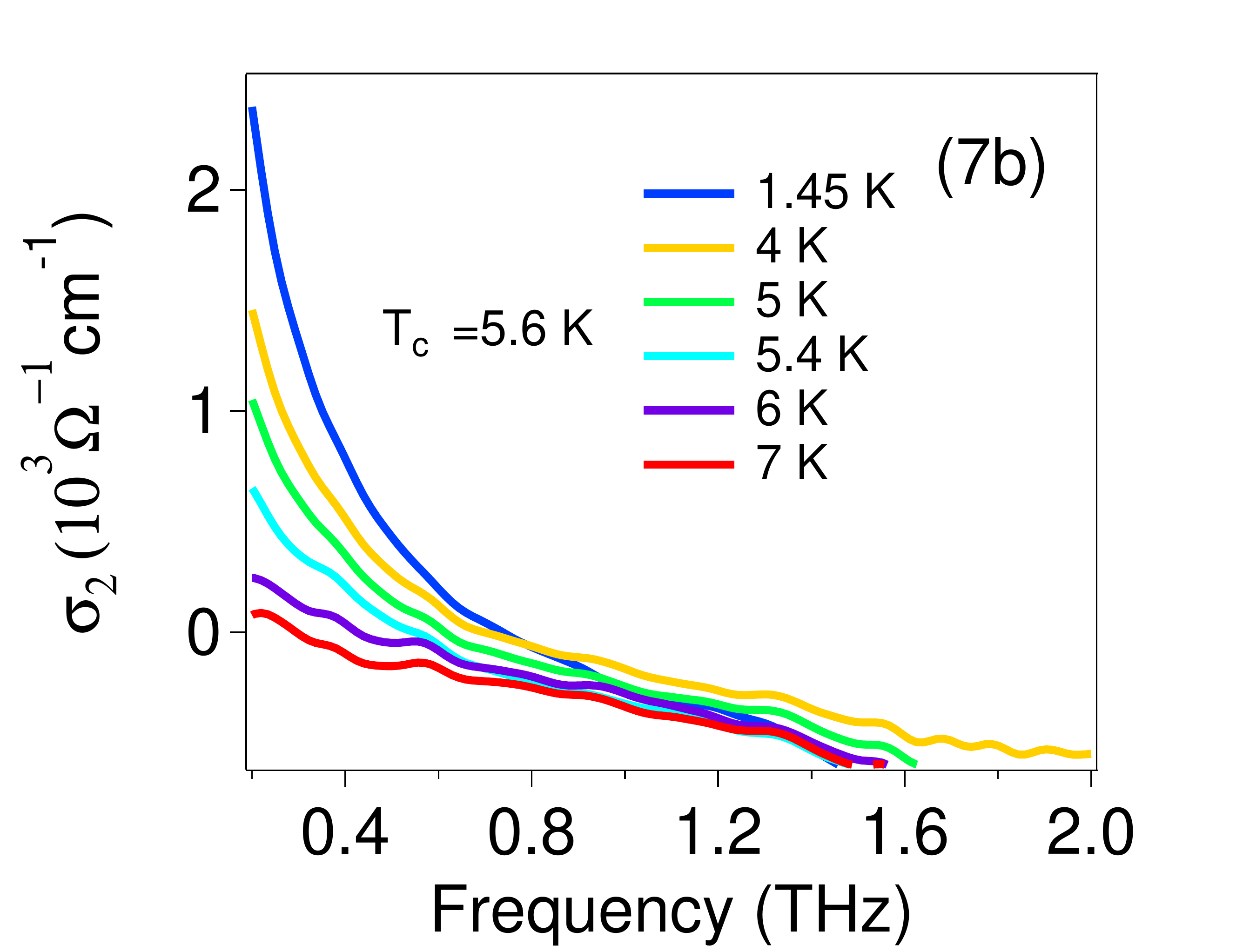}
\includegraphics[width=0.5\columnwidth]{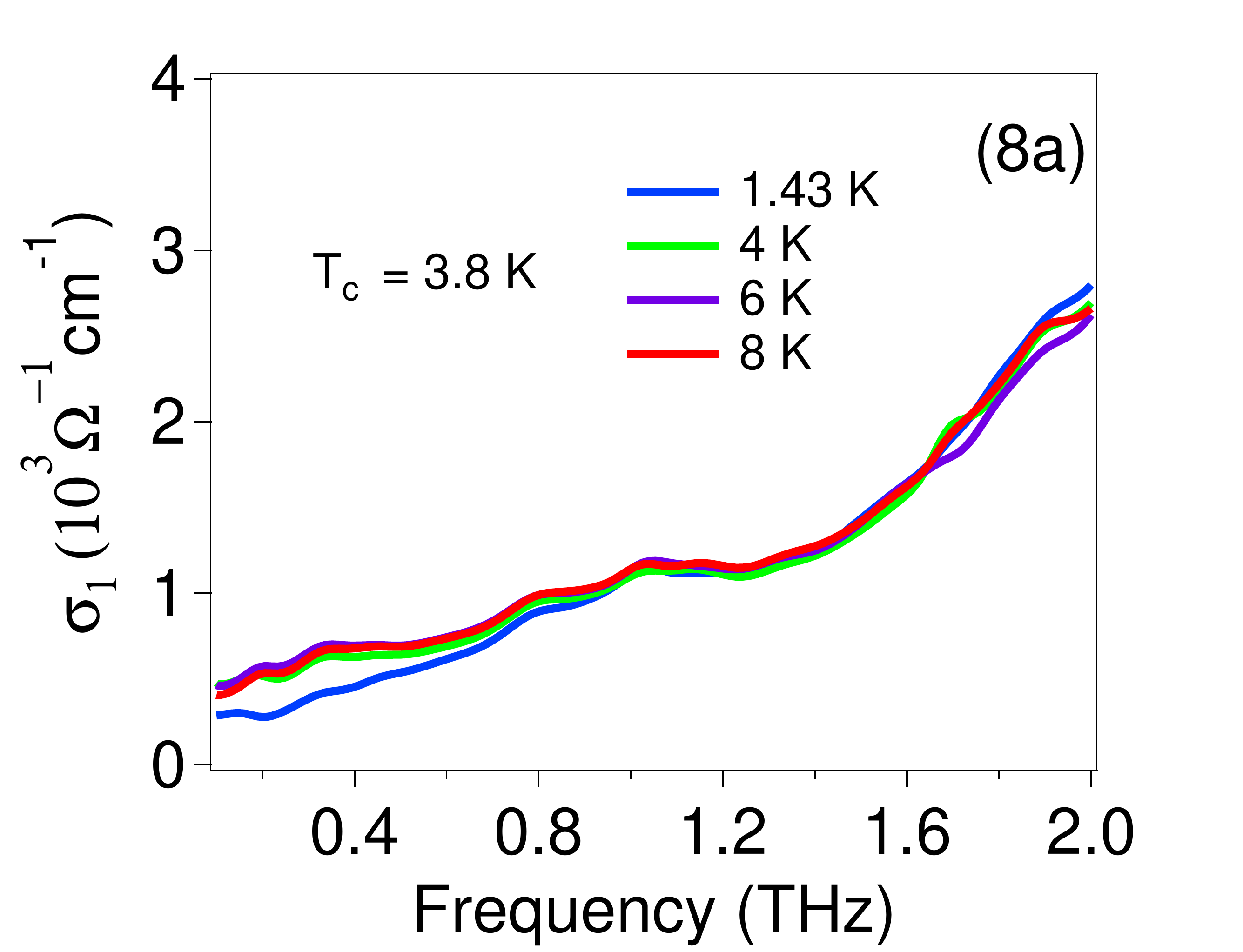}
\includegraphics[width=0.5\columnwidth]{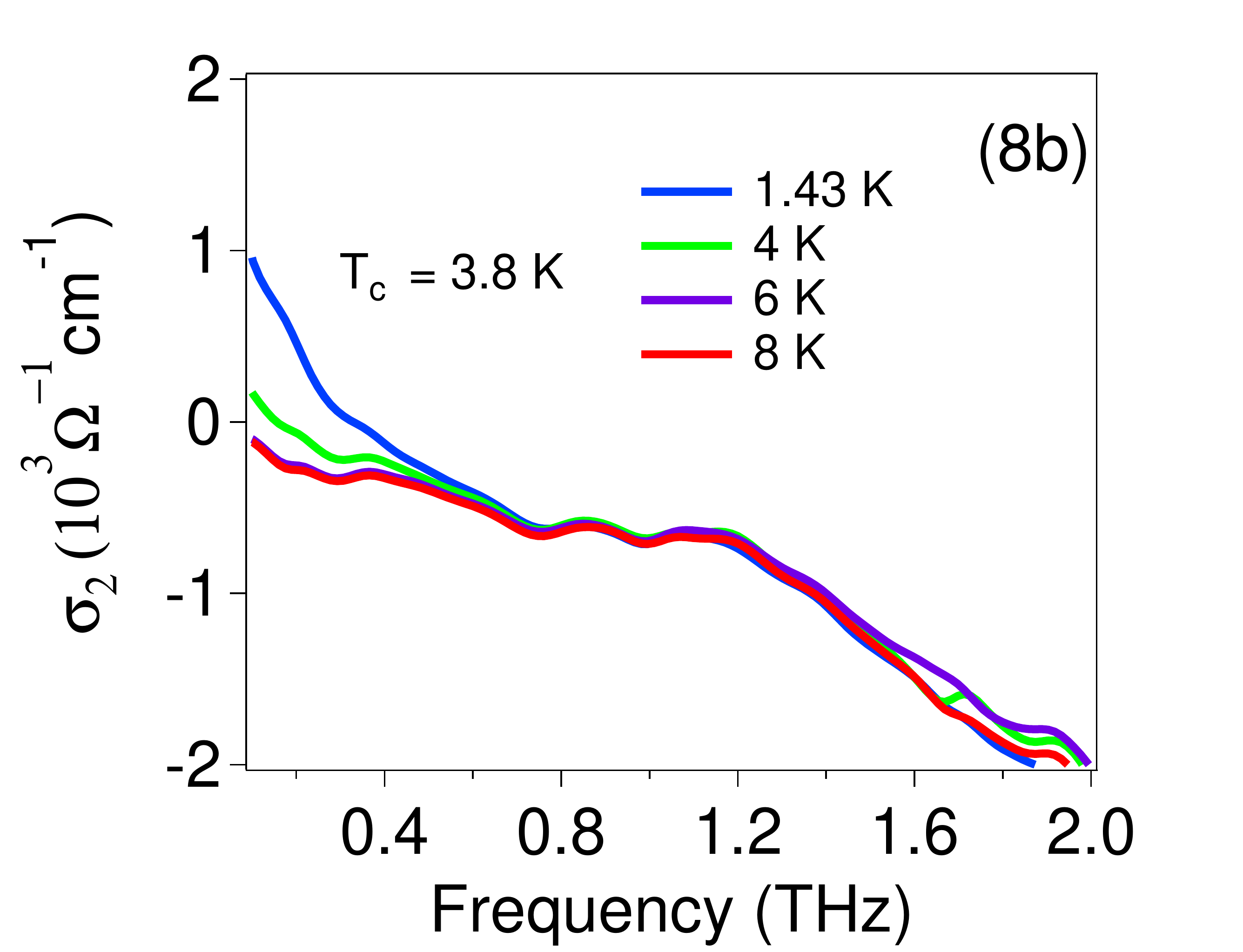}
\caption{(color online) Real and imaginary parts of temperature-dependent optical conductivity measurements for all samples in this work.   }
\end{figure}

\section{C. Extraction of Superconducting Gap $\Delta$ and De-pairing Factor $\eta$ from Planar Tunneling Experiments}
  
  The tunneling spectra for disordered NbN samples were measured by using planar superconductor/insulator/normal metal tunnel junctions comprising of reactively sputtered NbN, an insulating oxide layer and an Ag counter electrode. Details of fabrication and measurement can be found in Ref \cite{Raychaudhuri09a}. The spectra were fitted by using the theory of Larkin and Ovchinnikov and Feigel’man and Skvortsov \cite{AO,skvortsov12}. Ref \cite{AO} showed  that if the short-scale disorder in the form of a spatially varying BCS coupling constant is introduced, the effective pair-breaking is equivalent to that produced by magnetic impurities \cite{AG}. The best fit curves using this functional form are shown in Figure 2.

\begin{figure}[h]
\includegraphics[scale=0.7]{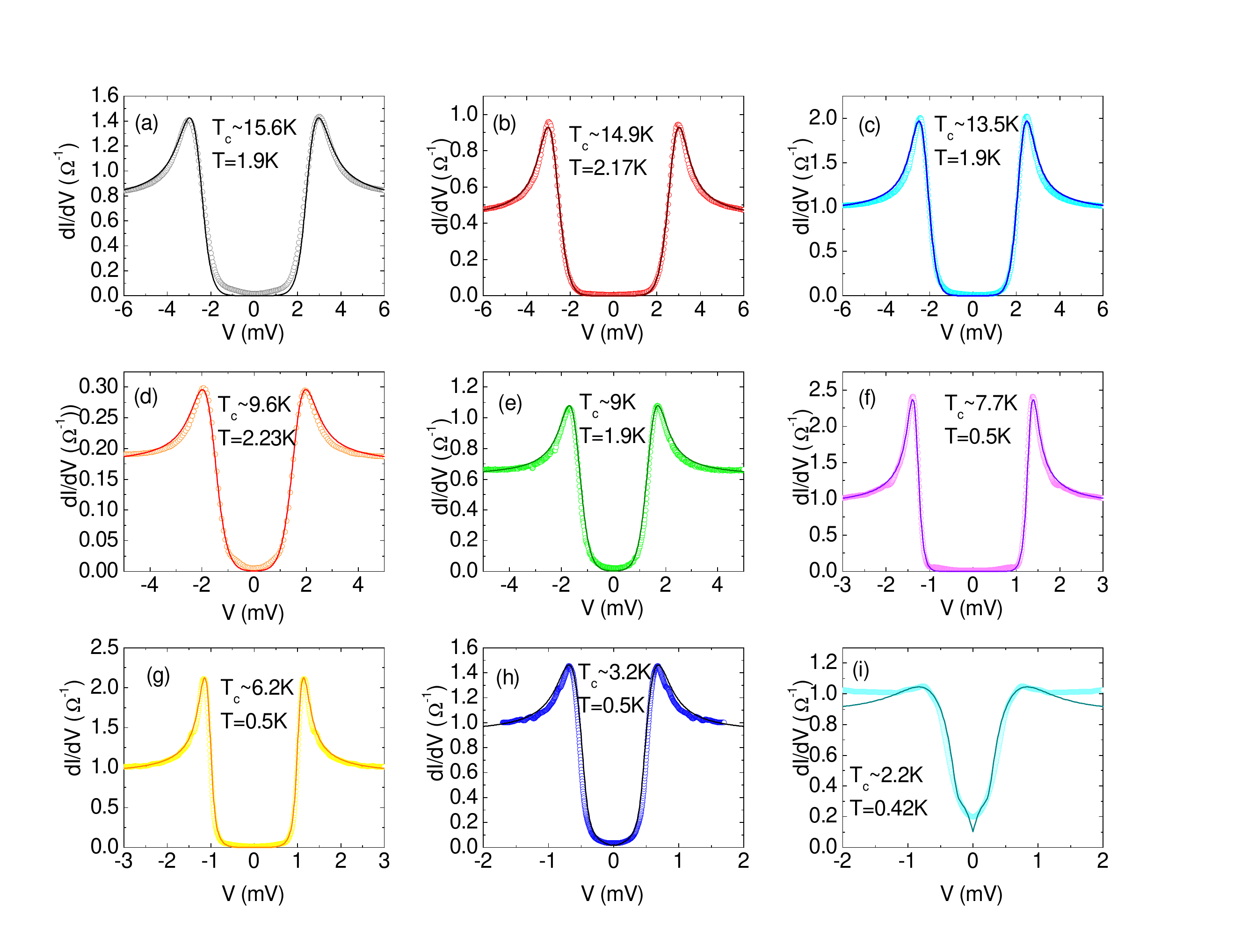}
\caption{(color online) Fitted tunneling spectra for samples with (a) $T_c\sim 15.6$ K, (b) $T_c \sim 14.9$ K, (c) $T_c\sim 13.5$ K, (d) $T_c\sim 9.6$ K, (e) $T_c\sim 9$ K, (f) $T_c\sim 7.7$ K, (g) $T_c\sim 6.2$ K, (h) $T_c\sim 3.2$ K, (i) $T_c\sim 2.2$ K}
\label{fits}
\end{figure}

\begin{figure}[h]
\includegraphics[scale=0.3]{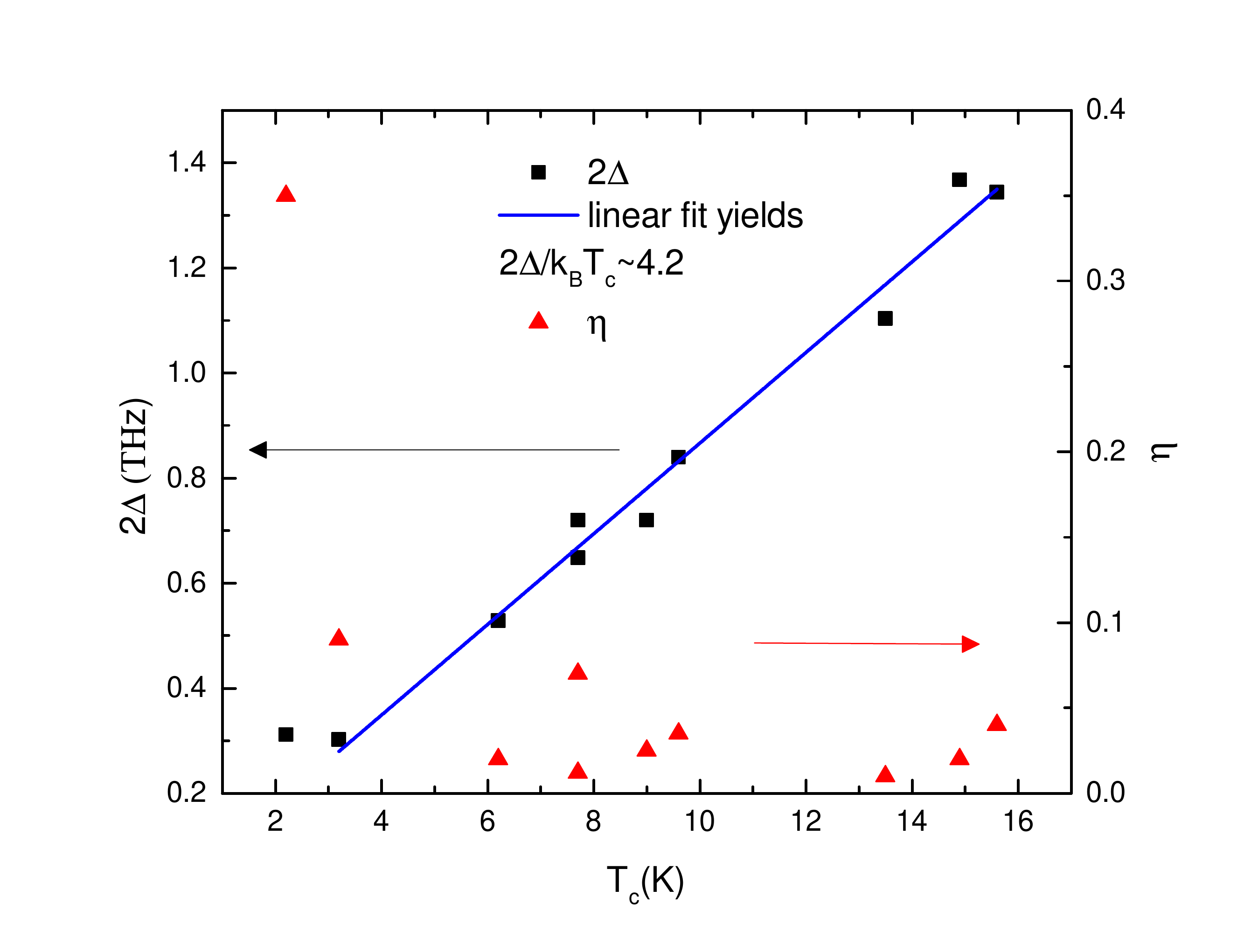}
\label{fits}
\caption{(color online) Fitted tunneling spectra for samples with $\Delta$ (left axis) and $\eta$ (right axis)values extracted from the fits of Figure 2 as functions of $T_c$}.
\end{figure}

\begin{figure}[H]
\includegraphics[width=0.5\columnwidth]{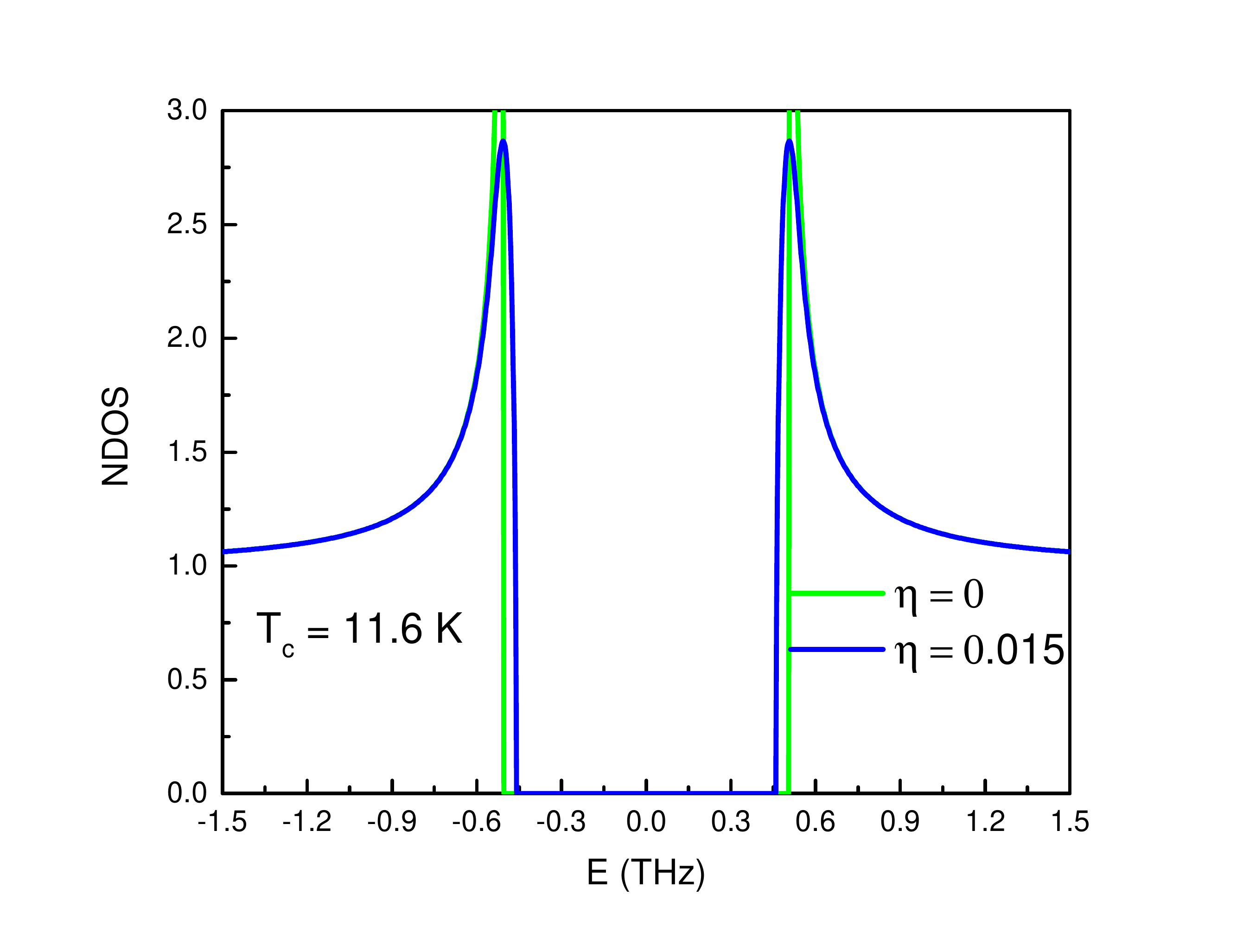}
\includegraphics[width=0.5\columnwidth]{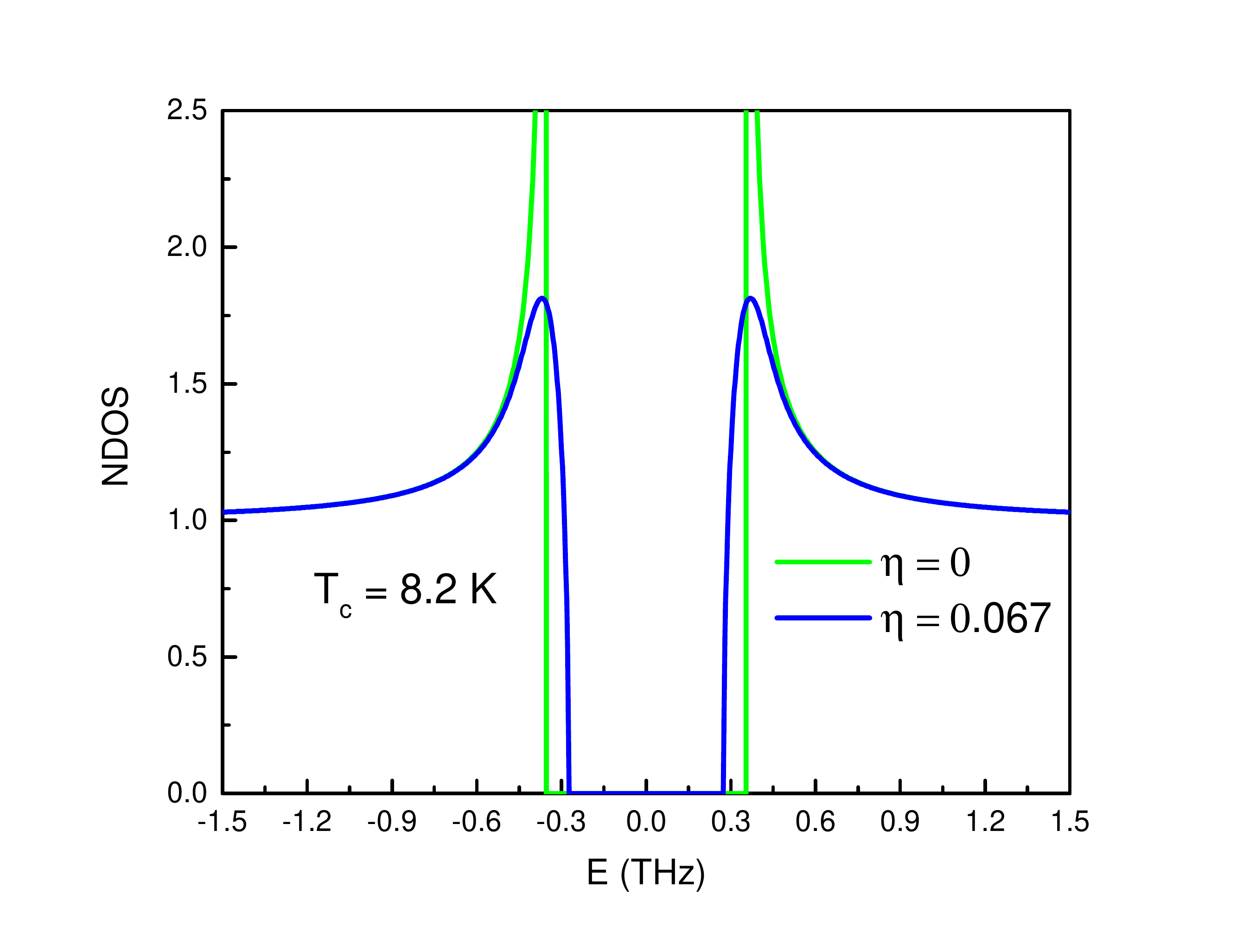}
\includegraphics[width=0.5\columnwidth]{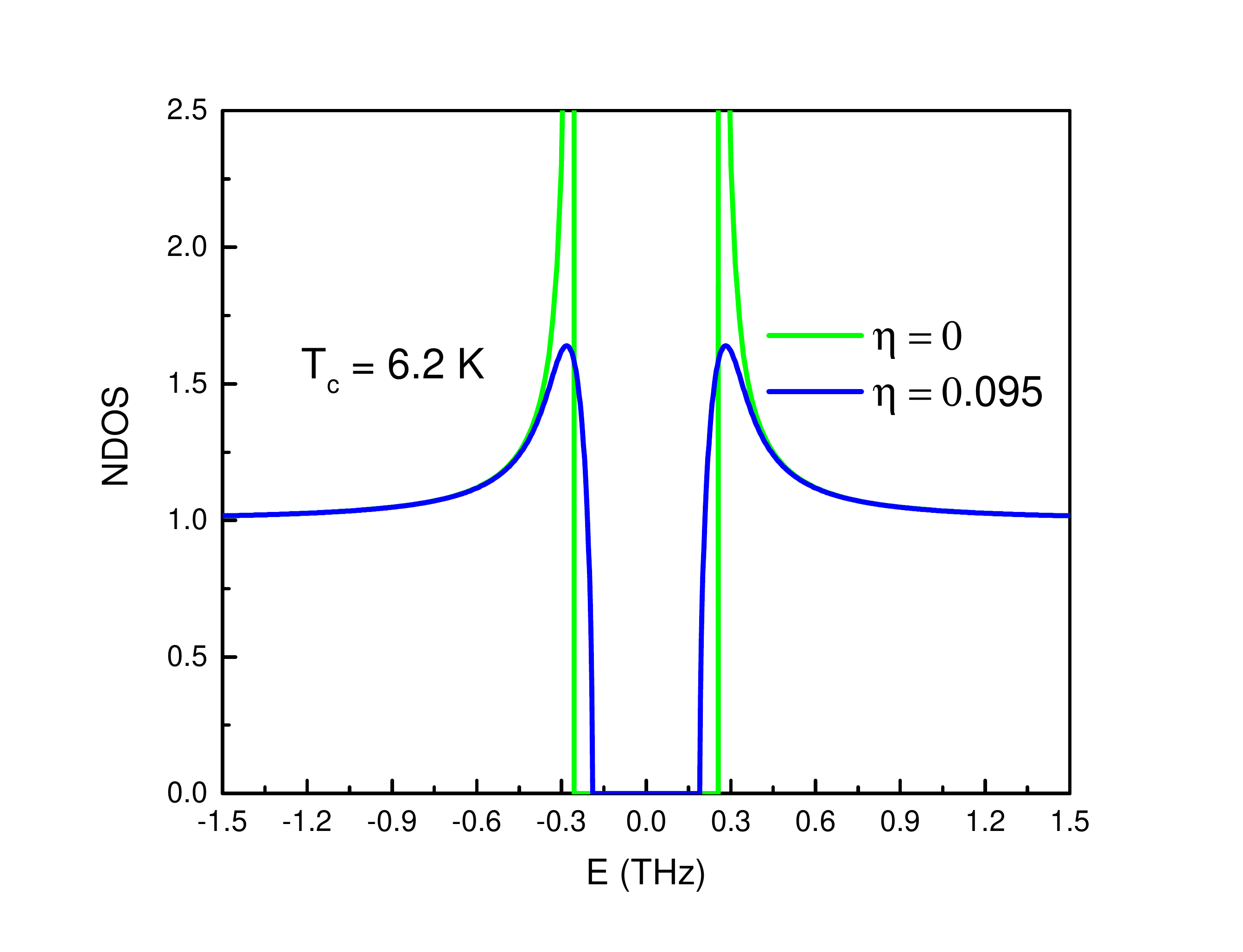}
\includegraphics[width=0.5\columnwidth]{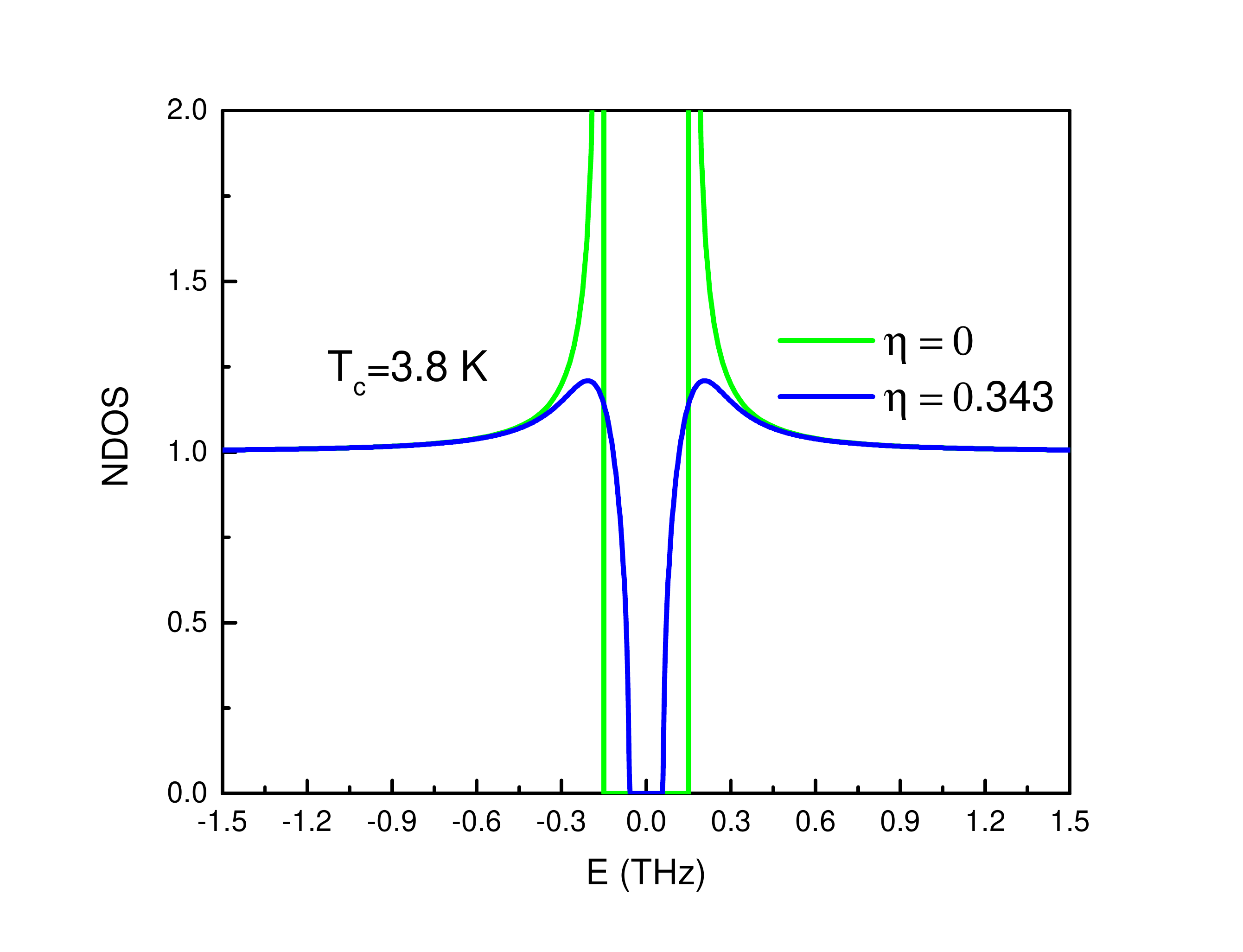}
\caption{(color online) Normalized density of states at T = 0 K predicted in the LO model. The parameters we use for simulations are the same as we simulate the normalized conductivity in the main text.    }
\end{figure}

To obtain these fits, the equation $u\left(1+\eta\dfrac{\sqrt{1-u^2}}{u^2-\epsilon_0^2}\right)=\dfrac{E}{\Delta}$ was first solved for $u$ \cite{Glazman}. Here $\eta$ is a phenomenological de-pairing factor that depends on the degree of disorder and $\epsilon_0$, which can be shown to determine the centre of the impurity band, was taken equal to unity. $\eta$ is a lower bound for the broadening parameter which arises from mesoscopic disorder. Other de-pairing factors not included in the theory could further increase its value.

The numerical solution obtained was used to generate the density of states (DOS) as $N(E)=N(0)Re\left(\dfrac{u}{u^2-1}\right)$. 
This resulted in a DOS with a hard gap renormalized to $E_g=(1-\eta^{2/3})^{3/2}\Delta$. 
The inclusion of electron-electron interaction effects modifies this DOS further to create sub-gap states \cite{AO,skvortsov12}, with $N(E)\propto \exp\left(\dfrac{E-E_g}{\Gamma_{tail}}\right)^{5/4}$ for $E<E_g$  in a 3D system. The DOS functions for $E>E_g$ and $E<E_g$ were pieced together to make a continuous function and then convoluted with the Fermi function to generate a fit for the observed conductance spectrum. $\Delta$ and $\eta$ were varied to obtain the best fit for the peaks, and $\Gamma_{tail}$ and the proportionality constant were adjusted to best fit the low-bias region. 

Figure 3 shows the resulting $\Delta$ (in THz units) and $\eta$ as a function of $T_c$. A linear fit to the data indicates $2\Delta/k_BT_c$ around 4.2, which is close to that reported in recent literatures \cite{Raychaudhuri09a,Chand12a,mintuthesis}. In very strongly disordered samples, the ratio becomes even larger as the gap persists above $T_c$ \cite{Raychaudhuri11}.

\section{D. Normalized density of states predicted by the Larkin-Ovchinnikov model in disordered superconductors}

As discussed in the main text, disorder smears the coherence peaks and transfers spectral weight into gap region. This feature can be captured by the model of Larkin and Ovchinnikov (LO). In Fig. 4, we show four simulations of the normalized density of states by using the LO model at typical values of $\eta$. The normalized density of states is determined by two factors: $\Delta$ and $\eta$. $\Delta$ sets the order parameter amplitude and is equivalent to the superconducting gap. $\eta$ is an effective parameter that captures depairing effects that are introduced by disorder. If $\eta$ = 0, the density of states simulated by the LO model automatically recovers the BCS prediction. As shown in the Fig. 4, the coherence peaks approach infinity with $\eta$ = 0.  With increasing $\eta$, one can see that even as the position of the coherence peaks (given approximately by the order parameter magnitude $\Delta$) does not change appreciably, substantial sub-gap tunneling conductance develops as the coherence peaks are smeared. As the disorder level increases, $\eta$ increases even more and additional density of states transfer into the region below $\Delta$.   Note that these simulations do not include the exponenential tail from localized states discussed in Sec. C.
\end{widetext}


\begin{thebibliography}{99}
\bibitem{Comment2}  In the theory of  Ref. \cite{AO} it is expected that localized states will give a small Lifshitz-like tail at energies below $E_g$.   We expect that these give a subdominant contribution to the electrodynamic response at low $\omega$ and are therefore neglected in our analysis.
\end{thebibliography}
\end{document}